\PassOptionsToPackage{table}{xcolor}
\documentclass[10pt,twocolumn,letterpaper]{article}

\usepackage{cvpr}      
\usepackage{multibib}
\makeatletter
\let\uemoSavedLN\@LN
\let\@LN\@gobbletwo
\newcites{app}{References}
\let\@LN\uemoSavedLN
\makeatother

\definecolor{cvprblue}{rgb}{0.21,0.49,0.74}
\usepackage[pagebackref,breaklinks,colorlinks,allcolors=cvprblue]{hyperref}
\usepackage{booktabs,multirow,graphicx}
\usepackage{array}

\usepackage{tcolorbox}
\tcbuselibrary{skins,breakable,listings}

\newtcblisting{uemoprompt}{
    enhanced,
    breakable,
    listing only,
    listing engine=listings,
    title={Unified Decision Prompt},
    colback=blue!5!white,
    colbacktitle=gray!8!white,
    colframe=blue!40!gray,
    coltitle=black,
    fonttitle=\bfseries,
    boxrule=0.6pt,
    arc=2mm,
    left=5pt,
    right=5pt,
    top=3pt,
    bottom=3pt,
    before skip=5pt,
    after skip=5pt,
    listing options={
        language={},
        basicstyle=\small\rmfamily,
        columns=fullflexible,
        keepspaces=true,
        breaklines=true,
        breakatwhitespace=true,
        showstringspaces=false,
        numbers=none,
        aboveskip=0pt,
        belowskip=0pt
    }
}

\def\confName{CVPR}
\def\confYear{2027}

\title{UEmo-D1: An Efficient and Calibrated Decision Model for Unified Multimodal Emotion Recognition}

\author{
Daiqing Wu\textsuperscript{1}\quad
Junqiao Wang\textsuperscript{1}\quad
Chengyang Fang\textsuperscript{2}\quad
Runhao Zeng\textsuperscript{3}\quad
Fei Ma\textsuperscript{4}\quad
Sicheng Zhao\textsuperscript{1}\\[3pt]
\textsuperscript{1}Tsinghua University\quad
\textsuperscript{2}Jiangxi University of Finance and Economics\\
\textsuperscript{3}Shenzhen MSU-BIT University\quad
\textsuperscript{4}Guangming Laboratory\\[-2pt]
{\ttfamily\small
wudaiqing@mail.tsinghua.edu.cn,\quad
schzhao@tsinghua.edu.cn
}
}

\begin{document}
\maketitle
\begin{abstract}
Multimodal Emotion Recognition (MER) seeks to identify emotions conveyed through or elicited by media comprising diverse modality compositions. Research in this field has largely evolved along two paradigms: task-specific models and approaches based on Multimodal Large Language Models (MLLMs). Despite their distinct strengths, neither fully reconciles three capabilities essential to human-like emotion understanding: cross-task generalization, efficient decision-making, and receptiveness to multiple plausible interpretations. Inspired by recent advances in decision models, we explore a unified framework that maps multimodal inputs, guided by natural language instructions, directly to distributions over candidate options, addressing these requirements through direct probabilistic decisions. We instantiate this framework in UEmo-D1, the first decision model for MER to our knowledge. To support joint training and systematic evaluation, we also curate UEmo-Data, comprising 977,270 training samples and 41 evaluation subsets across six MER tasks. Extensive experiments demonstrate UEmo-D1's combination of strong recognition performance, low inference latency, and well-calibrated confidence estimates. Complementary analyses further reveal how training strategies and data composition influence these capabilities, providing practical insights into unified MER modeling. Together, our findings establish decision models as a feasible direction toward human-level emotion recognition. Project page: \url{https://github.com/wdqqdw/UEmo-D1}.
\end{abstract}    
\begin{figure}[t]
    \centering
    \includegraphics[width=1\linewidth]{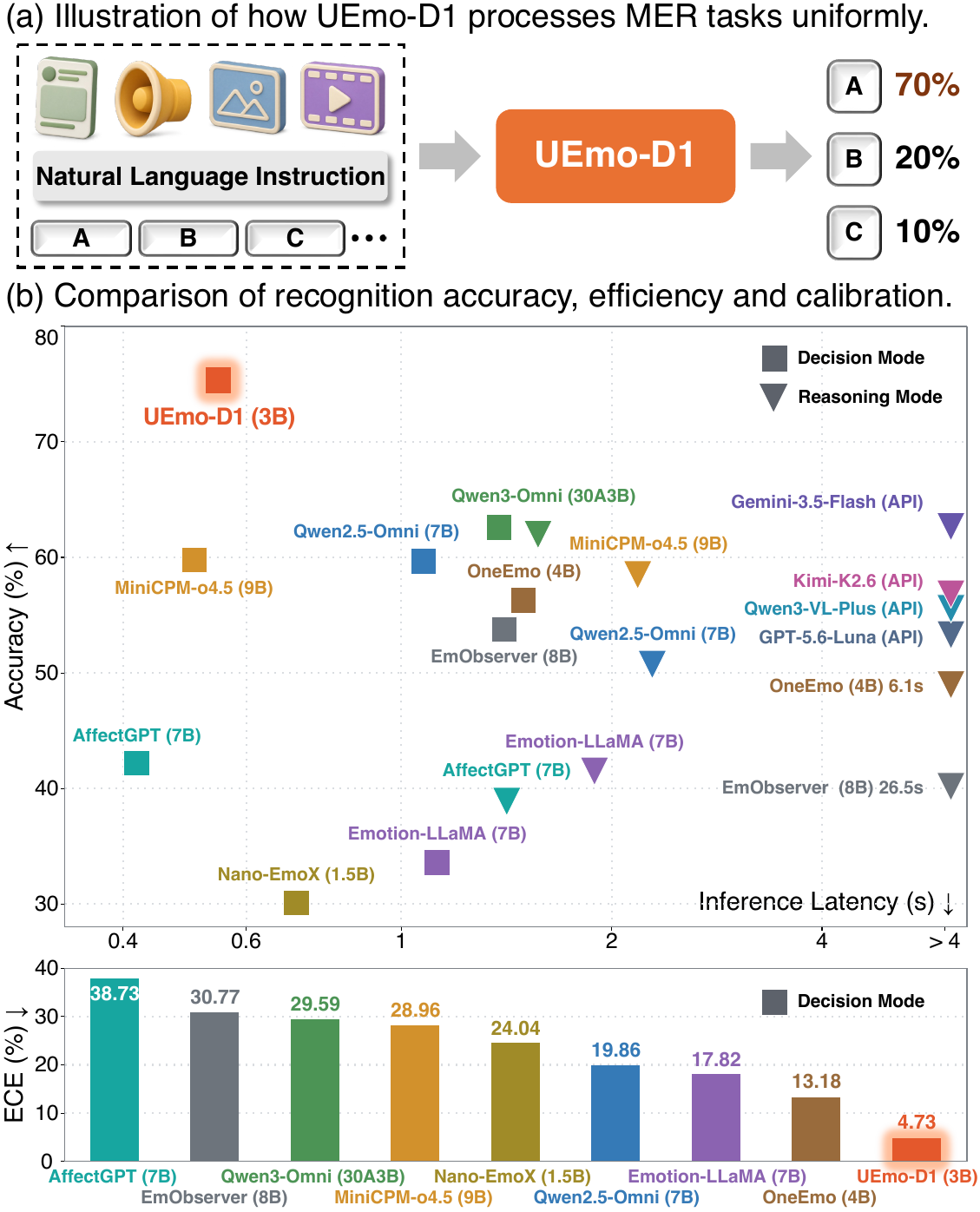}
    \vskip -0.10in
    \caption{Illustration of UEmo-D1's decision-making framework and comparison of its accuracy, efficiency, and calibration with advanced MLLMs, with performance averaged over 41 subsets spanning six MER tasks. Decision mode compares first-token option probabilities, whereas reasoning mode allows for deliberation. Inference latency is measured on two NVIDIA RTX 4080 GPUs.}
    \label{fig:1}
    \vskip -0.10in
\end{figure}

\section{Introduction}
\label{sec:intro}

Emotion plays an integral role in shaping human behavior and social interactions~\cite{keltner1999social}. Understanding how people express and respond to emotions is therefore important for effective communication~\cite{vankleef2009emotions}, with broader implications for building meaningful interpersonal relationships~\cite{lopes2003emotional} and monitoring mental health~\cite{dibeklioglu2015multimodal}. Multimodal Emotion Recognition (MER) seeks to equip machines with this ability by drawing on multimodal content to identify emotions directly expressed by people~\cite{poria2019meld}, conveyed through media~\cite{li2022clmlf}, or evoked by media~\cite{yang2023emoset}. Given its close connection to human experience and broad application potential, MER has attracted rapidly growing interest from both academia and industry~\cite{poria2017review}.

After decades of research and debate, MER has evolved into two broad paradigms with distinct characteristics: compact task-specific models~\cite{zhao2021emotion} and approaches based on multimodal large language models (MLLMs)~\cite{zhang2026merllm}. Task-specific models excel at learning mappings between predefined input and output spaces, typically through supervised training on dedicated datasets. Although computationally efficient, these models often generalize poorly beyond their training settings~\cite{tu2026satisfactory}, suffering substantial performance degradation when input modalities are disrupted~\cite{zhao2021mmin}, the target label space changes~\cite{lian2025ovmer}, or even the input domain shifts~\cite{gideon2021addog}. By contrast, MLLM-based MER reverses this trade-off, drawing on extensive world knowledge and strong instruction-following capabilities~\cite{touvron2023llama2} to achieve broader generalization at greater computational cost~\cite{xie2024emovit}. More recently, the field is increasingly embracing explicit affective reasoning before prediction to make emotion judgments more explainable and accurate~\cite{lan2025expllm,lian2023emer}.

Despite their respective strengths, the two paradigms face differing limitations in approaching human-level performance along three key dimensions. \textbf{First}, humans draw on partly shared cognitive representations~\cite{skerry2014common} to flexibly apply their emotional understanding across diverse contexts, from perceiving a speaker's emotions in face-to-face interactions to recognizing emotions evoked by media~\cite{cowen2019mapping}. Achieving comparable versatility remains challenging even for MLLM-based approaches, whose generalization typically operates under fixed task requirements rather than across different task demands~\cite{lian2026affectgptrl}. \textbf{Second}, human emotion perception is often intuitive and rapid, requiring little deliberate thought~\cite{tracy2008automaticity}. This resembles the direct inference of compact task-specific models, albeit without their generalization bottlenecks, and challenges the need for elaborate reasoning and its attendant computational overhead in MLLM-based approaches. \textbf{Third}, human emotion understanding inherently accommodates multiple plausible interpretations, reflecting a distribution of possibilities rather than a single categorical judgment~\cite{ong2015affective}. This flexibility facilitates meaningful interactions with individuals from diverse backgrounds and different expressive styles, a capacity that neither paradigm is sufficiently adept at capturing and reproducing.

The emerging paradigm of decision models, often termed System 1 models~\cite{almeida2026jev,geuter2026releasing,li2026llmasjev}, offers a promising avenue for bridging these gaps by combining flexible input processing with direct probabilistic decisions in place of explicit reasoning. Inspired by this paradigm, we introduce \textbf{UEmo-D1}, to our knowledge, the first decision model for unified MER. Guided by natural-language instructions and user-defined closed-set options, UEmo-D1 maps heterogeneous inputs, ranging from facial and vocal expressions to social media posts and audiovisual conversations, to task-specific decisions within a single framework, as illustrated in \cref{fig:1}(a). Each decision is accompanied by a probability distribution over the candidate options, obtained at a single decision-token position. To support this unified formulation with diverse supervision and comprehensive evaluation, we construct \textbf{UEmo-Data}, bringing together nearly one million training samples and 41 test subsets spanning diverse MER tasks, including facial expression recognition~\cite{wang2020scn}, conversational multimodal emotion recognition~\cite{hu2021mmgcn}, and image-text sentiment analysis~\cite{xu2017multisentinet}. As shown in \cref{fig:1}(b), extensive comparisons against leading emotion-oriented, open-source, and proprietary MLLMs demonstrate that UEmo-D1 achieves strong recognition performance across tasks while maintaining low inference latency and markedly better-calibrated predictions. By bringing these capabilities together, UEmo-D1 offers a unified, efficient, and calibrated alternative to the two prevailing paradigms, advancing decision modeling as a practical route toward more human-like emotion recognition. Collectively, our main contributions are threefold:

\begin{itemize}
    \item We introduce UEmo-D1, the first decision model for unified MER, bringing together versatile emotion recognition, low-latency inference, and well-calibrated confidence estimation within a single framework.
    
    \item We present UEmo-Data, a comprehensive resource that consolidates nearly one million heterogeneous samples from diverse MER sources into a unified interface, enabling joint learning and systematic evaluation.

    \item Extensive experiments establish UEmo-D1's strengths in fine-grained recognition and confidence calibration, while elucidating how training strategies and data recipes shape decision-model performance. These findings provide practical guidance for developing decision models as a unified alternative for MER.
\end{itemize}

\begin{figure*}[t]
    \centering
    \includegraphics[width=1\linewidth]{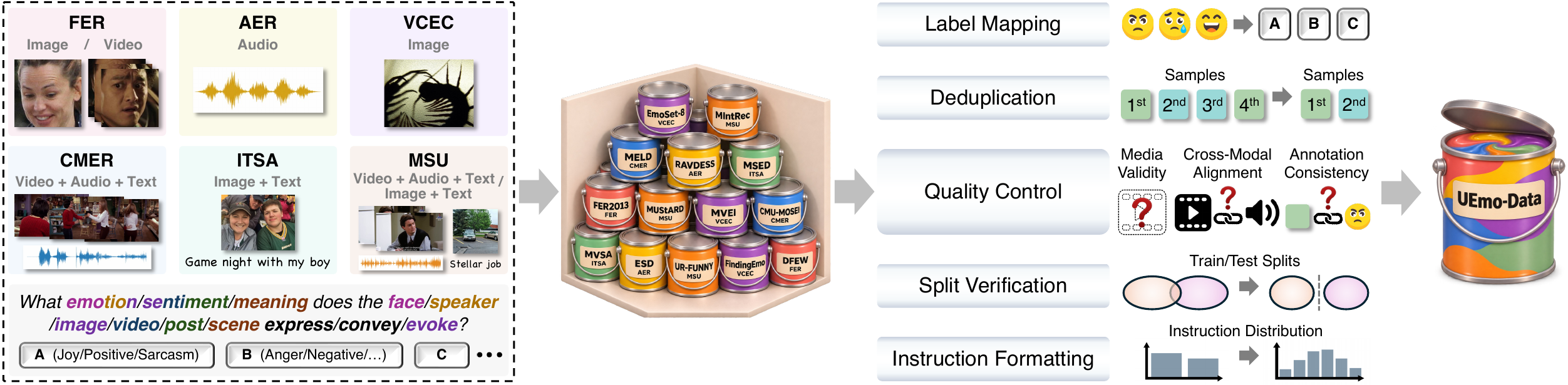}
    \vskip -0.10in
    \caption{Overview of the UEmo-Data construction pipeline. Heterogeneous datasets from diverse sources spanning six MER tasks are collected and processed through five stages: label mapping, deduplication, quality control, split verification, and instruction formatting. The processed data are then consolidated into UEmo-Data, a unified resource for MER training and evaluation.}
    \label{fig:2}
    \vskip -0.10in
\end{figure*}

\begin{figure*}[t]
    \centering
    \includegraphics[width=1\linewidth]{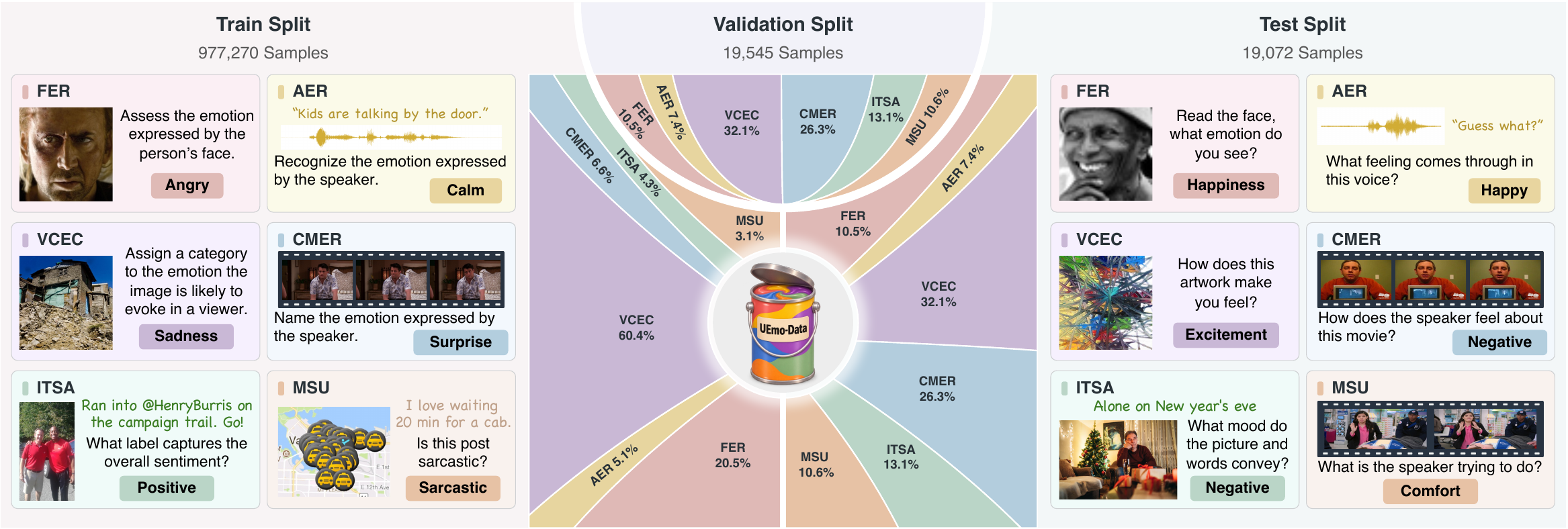}
    \vskip -0.10in
    \caption{Representative examples and task distributions of UEmo-Data. Reflecting the abundance of high-quality visual resources, VCEC and FER account for the largest shares of the 977,270 training samples, while the validation and test splits, comprising 19,545 and 19,072 samples respectively, maintain more balanced distributions across the six MER tasks.}
    \label{fig:3}
    \vskip -0.10in
\end{figure*}

\section{Related Work}
\paragraph{Multimodal Emotion Recognition.}
MER represents a broad research area in which studies adopt the same MER designation despite differences in recognition targets, input modalities, and task requirements. Representative tasks include recognizing speakers' emotions from audiovisual utterances~\cite{li2024corrkd}, identifying affect conveyed by image--text social media posts~\cite{kumar2021hybrid}, and interpreting emotional states alongside human behaviors in multimodal videos~\cite{zhao2025humanomni}. With the emergence of MLLM-based approaches, the scope of MER has expanded further. Their language-centered design makes text an integral input modality, allowing emotion recognition from vision or audio alone to fall within a common multimodal framework~\cite{zhang2026merllm}.

Early MLLM-based efforts, including EmoLA~\cite{li2024emola}, EmoVIT~\cite{xie2024emovit}, and Emotion-LLaMA~\cite{cheng2024emotionllama}, leverage natural-language instructions to bring previously separate recognition settings into shared models, enabling generalization across datasets and label taxonomies within their respective task domains. Building on this foundation, recent work such as AffectGPT~\cite{lian2025affectgpt} and EmoCaliber~\cite{wu2025emocaliber} places increasing emphasis on affective reasoning to improve explainability and recognition accuracy. Structured reasoning grounds emotional judgments in perceptual evidence, progressing from salient entities and their attributes to contextual relationships and, ultimately, emotional interpretations.

\subsection{Unified Modeling of MER Tasks}
Motivated by the interconnected nature of human affective processing, a newly emerging wave of research has focused on generalization across MER tasks. Nano-EmoX~\cite{huang2026nanoemox} integrates facial, vocal, and linguistic cues to jointly recognize emotions and sentiments, infer intentions, and explain emotional causes. OneEmo~\cite{huang2026oneemo} further incorporates social understanding, including humor and sarcasm, within a unified multitask framework. EmObserver~\cite{wu2026emobserver} explores a complementary direction, bringing diverse aspects of visual emotional comprehension into a common statement-judgement formulation. Other efforts extend unification beyond understanding to emotion-evoking image generation, as exemplified by UniEmo~\cite{zhu2026uniemo}. Despite their effectiveness, these approaches remain constrained in input-modality coverage, often rely on elaborate reasoning, and fail to adequately reflect the distributional nature of emotional interpretations, as quantitatively reflected in Fig.~\ref{fig:1}(b). In contrast, UEmo-D1 adopts an alternative decision-modeling paradigm, delivering clear and simultaneous gains in recognition performance, inference efficiency, and confidence calibration.
\begin{figure*}[t]
    \centering
    \includegraphics[width=1\linewidth]{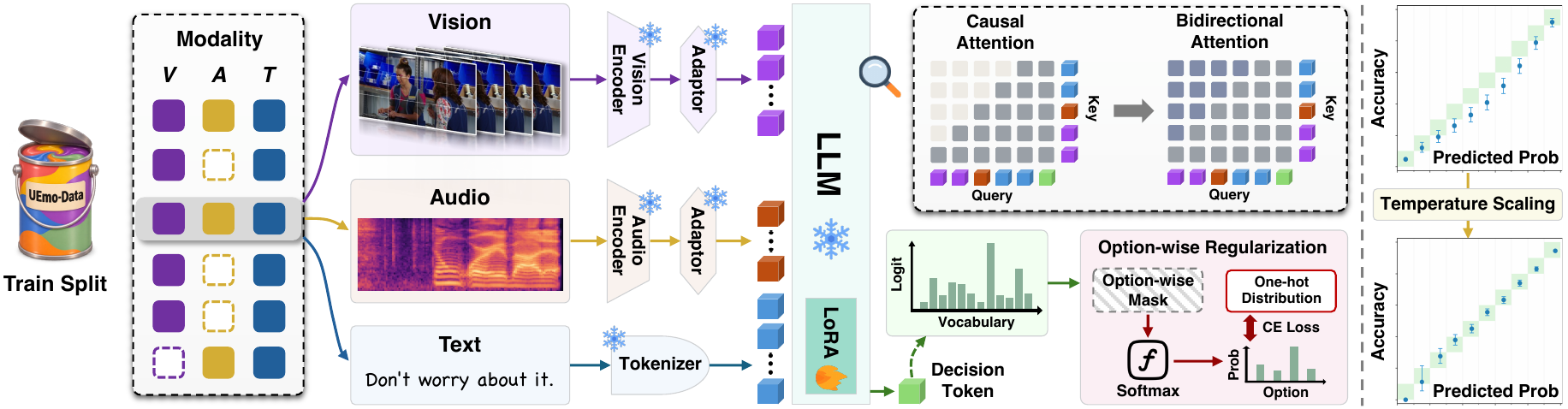}
    \vskip -0.10in
    \caption{Overview of the training and calibration pipeline for UEmo-D1. Starting from the UEmo-Data train split, samples with different modality combinations undergo unchanged modality-specific encoding before entering the language backbone, where causal attention is replaced with bidirectional attention to strengthen contextual modeling. At the output decision-token position, option-wise regularization then restricts softmax normalization to valid options for cross-entropy supervision. Throughout training, only LoRA parameters~\cite{hu2022lora} in the language backbone are updated, with all original model parameters frozen. After training, softmax temperature scaling is fitted on the validation split to yield the calibrated UEmo-D1.}
    \label{fig:4}
    \vskip -0.10in
\end{figure*}

\section{Construction of UEmo-Data}
\label{sec:3}
Data form the foundation of modern model development. To enable rigorous evaluation and informative supervision, we first construct UEmo-Data based on the task taxonomy established in a recent MER survey~\cite{zhang2026merllm}, extending it to include higher-level social understanding. The resulting resource covers six complementary MER tasks, with 41 evaluation subsets and 977,270 training samples. These tasks are defined as follows:
\textbf{(1)} \textit{Facial Expression Recognition} (FER)~\cite{wang2020scn,mollahosseini2019affectnet}, which identifies emotions expressed through facial configurations and movements in images or videos;
\textbf{(2)} \textit{Audio Emotion Recognition} (AER)~\cite{livingstone2018ravdess}, which recognizes emotions conveyed through vocal expressions and acoustic cues;
\textbf{(3)} \textit{Visual Content Emotion Comprehension} (VCEC)~\cite{xie2024emovit,yang2023emoset}, which infers the emotional responses that visual content is likely to evoke in viewers;
\textbf{(4)} \textit{Conversational Multimodal Emotion Recognition} (CMER)~\cite{hu2021mmgcn,poria2019meld}, which identifies a speaker's emotional state from the visual, acoustic, and linguistic cues accompanying an utterance;
\textbf{(5)} \textit{Image--Text Sentiment Analysis} (ITSA)~\cite{li2022clmlf,wu2025sentimental}, which determines the sentiment conveyed jointly by an image and its accompanying text;
and \textbf{(6)} \textit{Multimodal Social Understanding} (MSU)~\cite{zhang2022mintrec,castro2019sarcasm}, which integrates multimodal cues to infer communicative intent and recognize social meanings such as humor and sarcasm.

The construction of UEmo-Data is outlined in \cref{fig:2}. We begin by collecting datasets spanning diverse input modalities and task requirements, then process them through a five-stage pipeline summarized below, with further details provided in the appendix. \textbf{First}, with nearly all source datasets containing at most 26 categories, we adopt alphabetic option identifiers and map each dataset's predefined categories to consecutive letters in ascending order. \textbf{Second}, to account for shared multimodal sources across datasets, we perform content-based deduplication and adjust class distributions within individual datasets. \textbf{Third}, we conduct quality control by excluding samples with inaccessible media and correcting issues such as modality misalignment and annotation inconsistencies. \textbf{Fourth}, we coordinate training, validation, and test partitioning across the entire collection to prevent cross-dataset leakage while preserving original split assignments wherever specified. For datasets without predefined splits, we target 500 samples per test subset, subject to availability, and allocate additional available samples to training. We then reserve a validation set from the training pool, comparable in size to the test set. \textbf{Finally}, we systematically vary instruction phrasing, option ordering, and other aspects of task presentation to mitigate systematic biases and reduce reliance on shortcuts during both training and evaluation. This process yields UEmo-Data with three disjoint splits, whose representative examples and task distributions are demonstrated in \cref{fig:3}.
\begin{table*}[t]
\centering
\caption{
Recognition performance comparison of advanced open-source, proprietary, and emotion-oriented MLLMs on the test split of UEmo-Data. For models without AER support, uniform-random expected AER scores are used to compute the average scores. R and D denote reasoning and decision modes. \textbf{Bold} and \underline{underline} indicate the best and second-best results in each column.
}
\vskip -0.10in
\label{tab:1}

\setlength{\tabcolsep}{5pt}
\renewcommand{\arraystretch}{1.1}

\resizebox{0.94\textwidth}{!}{%
\begin{tabular}{@{}lc*{14}{c}@{}}
\toprule
\multirow{2}{*}{\textbf{Model}}
& \multirow{2}{*}{\textbf{Mode}}
& \multicolumn{2}{c}{FER (\%)}
& \multicolumn{2}{c}{AER (\%)}
& \multicolumn{2}{c}{VCEC (\%)}
& \multicolumn{2}{c}{CMER (\%)}
& \multicolumn{2}{c}{ITSA (\%)}
& \multicolumn{2}{c}{MSU (\%)}
& \multicolumn{2}{c}{Avg.} \\

\cmidrule(lr){3-4}
\cmidrule(lr){5-6}
\cmidrule(lr){7-8}
\cmidrule(lr){9-10}
\cmidrule(lr){11-12}
\cmidrule(lr){13-14}
\cmidrule(lr){15-16}

& & Acc & F1 & Acc & F1 & Acc & F1
  & Acc & F1 & Acc & F1 & Acc & F1 & Acc & F1 \\
\midrule

\rowcolor{black!8}
\multicolumn{16}{c}{Open-Source MLLMs} \\

Qwen2.5-Omni~\cite{xu2025qwen25omni}
{\footnotesize\textcolor{gray}{7B}}
& R
& 37.0 & 27.4
& 28.4 & 22.2
& 59.5 & 57.1
& 58.0 & 52.6
& 62.6 & 57.5
& 65.1 & 62.2
& 51.8 & 46.5 \\

Qwen2.5-Omni~\cite{xu2025qwen25omni}
{\footnotesize\textcolor{gray}{7B}}
& D
& 38.2 & 29.1
& \underline{54.4} & \underline{48.7}
& 61.0 & 58.8
& 70.3 & 55.6
& 64.1 & 57.6
& 67.1 & 63.4
& 59.2 & 52.2 \\

Qwen3-Omni~\cite{xu2025qwen3omni}
{\footnotesize\textcolor{gray}{30A3B}}
& R
& 51.4 & 38.3
& 42.1 & 34.7
& 63.4 & 61.2
& 71.4 & 57.0
& 63.9 & 59.1
& \underline{73.8} & \underline{72.8}
& 61.0 & 53.9 \\

Qwen3-Omni~\cite{xu2025qwen3omni}
{\footnotesize\textcolor{gray}{30A3B}}
& D
& 49.8 & 39.1
& 46.8 & 38.5
& 64.2 & 62.4
& \underline{71.7} & \underline{58.2}
& 63.2 & 55.8
& 72.6 & 71.0
& 61.4 & 54.2 \\

MiniCPM-o 4.5~\cite{cui2026minicpmo45}
{\footnotesize\textcolor{gray}{9B}}
& R
& 52.7 & 39.6
& 34.2 & 23.1
& 63.2 & 61.5
& 67.6 & 54.7
& 60.2 & 55.4
& 69.3 & 67.5
& 57.9 & 50.3 \\

MiniCPM-o 4.5~\cite{cui2026minicpmo45}
{\footnotesize\textcolor{gray}{9B}}
& D
& 54.3 & 41.4
& 37.9 & 27.4
& 63.0 & 61.3
& 69.1 & 53.0
& 60.4 & 55.2
& 71.0 & 69.5
& 59.3 & 51.3 \\

\rowcolor{black!8}
\multicolumn{16}{c}{Proprietary MLLMs} \\

Qwen3-VL-Plus~\cite{alibaba2026qwen3vlplus}
& R
& 56.0 & \underline{45.7}
& -- & --
& 63.2 & 61.5
& 68.5 & 54.7
& 63.5 & 58.4
& 67.2 & 64.9
& 56.0 & 50.3 \\

Kimi-K2.6~\cite{moonshot2026kimik26}
& R
& 52.1 & 40.6
& -- & --
& 63.9 & 62.0
& 71.5 & 56.3
& \underline{64.5} & \underline{61.3}
& 67.9 & 67.7
& 56.3 & 50.8 \\

Gemini-3.5-Flash~\cite{google2026gemini35flash}
& R
& 51.5 & 43.1
& 53.9 & 47.8
& 65.2 & 63.0
& 67.8 & 54.6
& 63.6 & 60.8
& 72.0 & 71.0
& \underline{62.3} & \underline{56.7} \\

GPT-5.6-Luna~\cite{openai2026gpt56luna}
& R
& 52.6 & 44.0
& -- & --
& 60.4 & 59.2
& 67.8 & 53.7
& 63.4 & 57.8
& 65.3 & 65.4
& 54.5 & 49.5 \\

\rowcolor{black!8}
\multicolumn{16}{c}{Emotion-Oriented MLLMs} \\

AffectGPT~\cite{lian2025affectgpt}
{\footnotesize\textcolor{gray}{7B}}
& R
& 11.0 & 8.3
& 20.5 & 13.1
& 29.7 & 23.6
& 54.8 & 43.5
& 56.8 & 49.7
& 55.6 & 55.6
& 38.1 & 32.3 \\

AffectGPT~\cite{lian2025affectgpt}
{\footnotesize\textcolor{gray}{7B}}
& D
& 15.2 & 9.4
& 23.0 & 13.5
& 34.3 & 28.0
& 62.4 & 50.6
& 58.1 & 50.8
& 61.9 & 58.7
& 42.5 & 35.2 \\

Emotion-LLaMA~\cite{cheng2024emotionllama}
{\footnotesize\textcolor{gray}{7B}}
& R
& 36.0 & 24.5
& 39.0 & 32.9
& 50.3 & 43.4
& 27.4 & 24.5
& 60.0 & 51.7
& 39.2 & 29.9
& 42.0 & 34.5 \\

Emotion-LLaMA~\cite{cheng2024emotionllama}
{\footnotesize\textcolor{gray}{7B}}
& D
& 39.6 & 23.2
& 21.4 & 16.7
& 41.5 & 37.2
& 21.7 & 18.5
& 50.0 & 42.8
& 34.7 & 24.9
& 34.8 & 27.2 \\

Nano-EmoX~\cite{huang2026nanoemox}
{\footnotesize\textcolor{gray}{1.5B}}
& R
& 0.5 & 0.7
& 16.5 & 14.3
& 3.7 & 5.3
& 16.4 & 17.3
& 37.2 & 36.6
& 5.2 & 9.1
& 13.3 & 13.9 \\

Nano-EmoX~\cite{huang2026nanoemox}
{\footnotesize\textcolor{gray}{1.5B}}
& D
& 14.4 & 6.5
& 19.0 & 12.4
& 32.4 & 27.4
& 54.6 & 35.1
& 23.9 & 18.8
& 38.6 & 30.1
& 30.5 & 21.7 \\

EmObserver~\cite{wu2026emobserver}
{\footnotesize\textcolor{gray}{8B}}
& R
& 32.0 & 27.8
& -- & --
& 64.2 & 64.5
& 60.0 & 38.6
& 55.2 & 50.0
& 12.0 & 11.9
& 40.2 & 34.9 \\

EmObserver~\cite{wu2026emobserver}
{\footnotesize\textcolor{gray}{8B}}
& D
& 52.6 & 39.4
& -- & --
& \underline{68.1} & \underline{66.3}
& 66.5 & 46.7
& 61.0 & 53.4
& 59.7 & 56.0
& 54.3 & 46.4 \\

OneEmo~\cite{huang2026oneemo}
{\footnotesize\textcolor{gray}{4B}}
& R
& 39.5 & 33.8
& 25.3 & 8.0
& 50.9 & 50.1
& 61.6 & 45.9
& 59.6 & 54.6
& 60.4 & 59.5
& 49.6 & 42.0 \\

OneEmo~\cite{huang2026oneemo}
{\footnotesize\textcolor{gray}{4B}}
& D
& \underline{56.2} & 40.9
& 27.5 & 7.5
& 57.2 & 55.3
& 67.3 & 45.4
& 62.6 & 52.3
& 60.9 & 57.7
& 55.3 & 43.2 \\

\rowcolor{blue!10}

\textbf{UEmo-D1 (Ours)}
{\footnotesize\textcolor{gray}{3B}}
& D
& \textbf{66.8} & \textbf{54.1}
& \textbf{72.8} & \textbf{71.2}
& \textbf{72.9} & \textbf{70.6}
& \textbf{77.6} & \textbf{62.7}
& \textbf{81.8} & \textbf{77.6}
& \textbf{81.5} & \textbf{80.7}
& \textbf{75.6} & \textbf{69.5} \\

\bottomrule
\end{tabular}%
}
\vskip -0.10in
\end{table*}

\section{Acquisition of UEmo-D1}
Consistent with recent studies on decision-model training~\cite{geuter2026releasing,li2026llmasjev}, our explorations reveal that dedicated training strategies struggle to accommodate heterogeneous data while maintaining reliable confidence estimates. Having dispensed with explicit reasoning, we also forgo an additional reinforcement learning stage given the risk of overconfidence associated with GRPO~\cite{shao2024deepseekmath,bereket2025uncalibrated,wu2025emocaliber}, opting instead for cross-entropy-based supervised fine-tuning (SFT) with three simple yet effective modifications.

\Cref{fig:4} illustrates the overall pipeline. We initialize our MLLM, denoted by $\pi_\theta$, from Qwen2.5-Omni-3B~\cite{xu2025qwen25omni} for its balance between efficiency and accuracy. Its native support for heterogeneous inputs allows missing visual or audio modalities to be omitted, while preserving the original encoding procedures for the available modalities. Within the language backbone, our first modification replaces the causal attention used for next-token pretraining with BERT-style~\cite{devlin2019bert} \textbf{Bidirectional Attention} over the input tokens during both training and inference. This enables each input token to attend to all others at every layer, strengthening contextual modeling within and across modalities. We denote the adapted model by $\pi_\theta^{*}$.

Given an input $x$ containing the available multimodal content, an instruction, and candidate options, $\pi_\theta^{*}$ predicts a single decision token $d$ without intermediate reasoning or subsequent autoregressive generation:
\begin{equation}
    d = \arg\max_{y \in \mathcal{C}}
    \pi_\theta^{*}(y|x),
\end{equation}
where $\mathcal{C}$ contains the valid option-letter tokens. Under standard training, the model produces logits $z_v$ over the vocabulary $\mathcal{V}$ at the decision position, followed by full-vocabulary softmax normalization and cross-entropy supervision. In place of this standard formulation, our second modification introduces \textbf{Option-wise Regularization} to focus learning on the candidate options. It confines probability normalization and supervision to the valid option set $\mathcal{C}$, with the resulting probability assigned to token $v$ defined as:
\begin{equation}
    p_v(\tau)
    =
    \frac{
        \mathcal{I}_{v \in \mathcal{C}}\exp(z_v/\tau)
    }{
        \sum_{u \in \mathcal{C}}\exp(z_u/\tau)
    },
\end{equation}
where $\mathcal{I}$ is an indicator function and $\tau$ is the temperature. With $y^{*}$ denoting the ground-truth option token, the training objective is:
\begin{equation}
    \mathcal{L}_{\mathrm{SFT}}
    =
    -\sum_{v \in \mathcal{C}}
    \mathcal{I}_{v=y^{*}}\log p_v(\tau=1).
\end{equation}

Cross-entropy training provides a principled foundation for calibrated probability estimation~\cite{gneiting2007strictly}. Building on the learned decision probabilities, our third modification refines confidence through post-training \textbf{Temperature Scaling} without updating model parameters. We fit the softmax temperature on a held-out validation set $\mathcal{D}_{\mathrm{val}}$ according to the expected calibration error (ECE) \cite{naeini2015obtaining}:
\begin{equation}
    \tau^{*}
    =
    \arg\min_{\tau>0}
        \sum_{b=1}^{B}
        \frac{|\mathcal{B}_b|}{|\mathcal{D}_{\mathrm{val}}|}
        \left|
            \operatorname{acc}(\mathcal{B}_b)
            -
            \operatorname{conf}_{\tau}(\mathcal{B}_b)
        \right|.
\end{equation}
Here, $\mathcal{B}_b$ denotes the $b$-th confidence bin based on $p_v(1)$, while $\operatorname{acc}$ and $\operatorname{conf}_{\tau}$ denote the empirical accuracy and mean confidence within each bin. At inference, we use $p_v(\tau^{*})$ as the calibrated option distribution.
\begin{figure*}[t]
    \centering
    \includegraphics[width=1\linewidth]{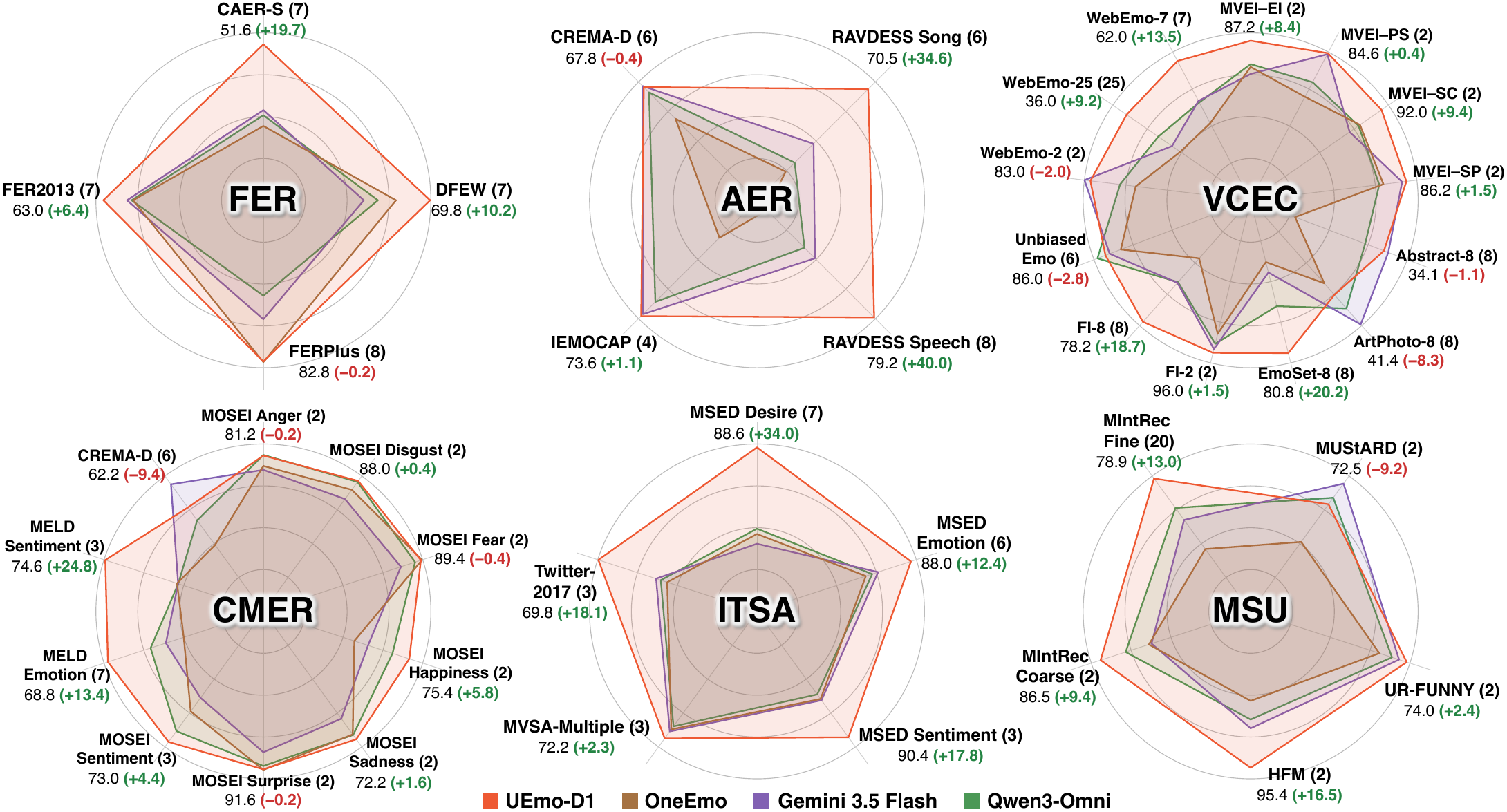}
    \vskip -0.10in
    \caption{Accuracy comparison of leading MLLMs on all 41 subsets spanning six MER tasks. Each subset is annotated with its number of candidate options, UEmo-D1's accuracy (\%), and its margin over the best-performing competing model in percentage points.}
    \label{fig:5}
    \vskip -0.10in
\end{figure*}

\begin{table*}[t]
\centering
\caption{
Calibration comparison of advanced MLLMs on the test split of UEmo-Data, all evaluated in decision mode. ECE and Brier scores are multiplied by 100. Unavailable AER results are excluded when computing the averages.}
\vskip -0.10in
\label{tab:2}

\setlength{\tabcolsep}{6pt}
\renewcommand{\arraystretch}{1.1}

\resizebox{0.96\textwidth}{!}{%
\begin{tabular}{@{}l*{14}{c}@{}}
\toprule
\multirow{2}{*}{\textbf{Model}}
& \multicolumn{2}{c}{FER}
& \multicolumn{2}{c}{AER}
& \multicolumn{2}{c}{VCEC}
& \multicolumn{2}{c}{CMER}
& \multicolumn{2}{c}{ITSA}
& \multicolumn{2}{c}{MSU}
& \multicolumn{2}{c}{Avg.} \\

\cmidrule(lr){2-3}
\cmidrule(lr){4-5}
\cmidrule(lr){6-7}
\cmidrule(lr){8-9}
\cmidrule(lr){10-11}
\cmidrule(lr){12-13}
\cmidrule(lr){14-15}

& ECE & Brier & ECE & Brier & ECE & Brier
& ECE & Brier & ECE & Brier & ECE & Brier
& ECE & Brier \\
\midrule

\rowcolor{black!8}
\multicolumn{15}{c}{Open-Source MLLMs} \\

Qwen2.5-Omni~\cite{xu2025qwen25omni}
{\footnotesize\textcolor{gray}{7B}}
& 21.9 & 80.1
& 17.2 & \underline{61.4}
& 24.3 & 60.3
& \underline{12.4} & \underline{41.4}
& 26.1 & 58.8
& 17.2 & 48.9
& 19.9 & 58.5 \\

Qwen3-Omni~\cite{xu2025qwen3omni}
{\footnotesize\textcolor{gray}{30A3B}}
& 30.3 & 77.1
& 45.5 & 95.9
& 28.8 & 62.0
& 21.3 & 47.6
& 32.8 & 67.8
& 18.9 & 44.3
& 29.6 & 65.8 \\

MiniCPM-o 4.5~\cite{cui2026minicpmo45}
{\footnotesize\textcolor{gray}{9B}}
& 27.8 & 70.5
& 47.1 & 102.6
& 27.0 & 61.8
& 22.6 & 51.2
& 33.8 & 70.6
& 15.5 & \underline{42.5}
& 29.0 & 66.5 \\

\rowcolor{black!8}
\multicolumn{15}{c}{Emotion-Oriented MLLMs} \\

AffectGPT~\cite{lian2025affectgpt}
{\footnotesize\textcolor{gray}{7B}}
& 52.9 & 127.0
& 46.2 & 113.3
& 47.1 & 106.6
& 23.1 & 60.2
& 35.8 & 75.2
& 27.3 & 62.7
& 38.7 & 90.8 \\

Emotion-LLaMA~\cite{cheng2024emotionllama}
{\footnotesize\textcolor{gray}{7B}}
& 14.1 & 65.3
& \underline{12.9} & 63.8
& \textbf{8.3} & \underline{41.2}
& 32.7 & 65.9
& \underline{8.9} & \underline{36.0}
& 29.6 & 62.5
& 17.8 & \underline{55.8} \\

Nano-EmoX~\cite{huang2026nanoemox}
{\footnotesize\textcolor{gray}{1.5B}}
& 25.6 & 94.7
& 28.3 & 95.1
& 19.1 & 78.1
& 19.1 & 59.0
& 26.3 & 89.0
& 25.8 & 73.7
& 24.0 & 81.6 \\

EmObserver~\cite{wu2026emobserver}
{\footnotesize\textcolor{gray}{8B}}
& 38.0 & 82.1
& -- & --
& 24.8 & 55.1
& 27.4 & 58.6
& 35.6 & 73.2
& 28.2 & 65.0
& 30.8 & 66.8 \\

OneEmo~\cite{huang2026oneemo}
{\footnotesize\textcolor{gray}{4B}}
& \underline{11.4} & \underline{57.0}
& 19.1 & 87.2
& 14.1 & 54.7
& 13.0 & 46.2
& 13.5 & 51.3
& \underline{8.0} & 49.0
& \underline{13.2} & 57.6 \\

\rowcolor{blue!10}
\textbf{UEmo-D1 (Ours)}
{\footnotesize\textcolor{gray}{3B}}
& \textbf{3.9} & \textbf{45.0}
& \textbf{4.9} & \textbf{36.2}
& \underline{8.6} & \textbf{37.7}
& \textbf{4.1} & \textbf{31.6}
& \textbf{2.8} & \textbf{24.2}
& \textbf{4.1} & \textbf{26.2}
& \textbf{4.7} & \textbf{33.5} \\

\bottomrule
\end{tabular}%
}
\vskip -0.10in
\end{table*}

\definecolor{deltagreen}{RGB}{35,132,67}
\definecolor{deltared}{RGB}{203,48,48}
\definecolor{oursbg}{RGB}{232,229,250}
\definecolor{basebg}{HTML}{E4EEDC}

\newcommand{\gplus}[1]{\,{\scriptsize\textcolor{deltagreen}{(+#1)}}}
\newcommand{\rminus}[1]{\,{\scriptsize\textcolor{deltared}{($-$#1)}}}
\newcommand{\rplus}[1]{\,{\scriptsize\textcolor{deltared}{(+#1)}}}
\newcommand{\gminus}[1]{\,{\scriptsize\textcolor{deltagreen}{($-$#1)}}}
\newcommand{\dzero}{\,{\scriptsize\textcolor{gray}{(0.00)}}}

\begin{table*}[t]
\centering
\caption{
Contributions of different task families to model training.
Row \#1 reports the Qwen2.5-Omni-3B baseline, with changes in rows \#2--\#7 measured relative to it.
Row \#8 reports UEmo-D1, with changes in rows \#9--\#14 measured relative to it.
}
\vskip -0.10in
\label{tab:4}
\setlength{\tabcolsep}{4pt}
\renewcommand{\arraystretch}{1.15}
\small
\resizebox{\textwidth}{!}{%
\begin{tabular}{@{}c
*{6}{>{\centering\arraybackslash}p{0.87cm}}
*{7}{>{\centering\arraybackslash}p{1.32cm}}
@{}}
\toprule
\multirow{2}{*}{Index}
& \multicolumn{6}{c}{Training Tasks}
& \multicolumn{7}{c}{Test Accuracy (\%) $\uparrow$} \\
\cmidrule(lr){2-7}
\cmidrule(lr){8-14}
& FER & AER & VCEC & CMER & ITSA & MSU
& FER & AER & VCEC & CMER & ITSA & MSU & Avg. \\
\midrule

\rowcolor{basebg}
\#1 & -- & -- & -- & -- & -- & --
& \textbf{34.5} & \textbf{47.0} & \textbf{60.9}
& \textbf{64.7} & \textbf{59.5} & \textbf{57.9}
& \textbf{54.1} \\

\#2 & \cellcolor{gray!12}\checkmark & -- & -- & -- & -- & --
& \cellcolor{gray!12}65.4\gplus{30.9}
& 45.2\rminus{1.8}
& 58.4\rminus{2.5}
& 57.2\rminus{7.5}
& 61.2\gplus{1.7}
& 58.9\gplus{1.0}
& 57.7\gplus{3.6} \\

\#3 & -- & \cellcolor{gray!12}\checkmark & -- & -- & -- & --
& 38.4\gplus{3.9}
& \cellcolor{gray!12}73.7\gplus{26.7}
& 57.0\rminus{3.9}
& 64.1\rminus{0.6}
& 59.8\gplus{0.3}
& 59.4\gplus{1.5}
& 58.7\gplus{4.6} \\

\#4 & -- & -- & \cellcolor{gray!12}\checkmark & -- & -- & --
& 29.3\rminus{5.2}
& 50.7\gplus{3.7}
& \cellcolor{gray!12}72.7\gplus{11.8}
& 63.1\rminus{1.6}
& 59.8\gplus{0.3}
& 59.3\gplus{1.4}
& 55.8\gplus{1.7} \\

\#5 & -- & -- & -- & \cellcolor{gray!12}\checkmark & -- & --
& 37.4\gplus{2.9}
& 37.4\rminus{9.6}
& 60.0\rminus{0.9}
& \cellcolor{gray!12}77.0\gplus{12.3}
& 62.6\gplus{3.1}
& 60.7\gplus{2.8}
& 55.9\gplus{1.8} \\

\#6 & -- & -- & -- & -- & \cellcolor{gray!12}\checkmark & --
& 39.5\gplus{5.0}
& 41.5\rminus{5.5}
& 59.8\rminus{1.1}
& 67.8\gplus{3.1}
& \cellcolor{gray!12}82.4\gplus{22.9}
& 56.6\rminus{1.3}
& 57.9\gplus{3.8} \\

\#7 & -- & -- & -- & -- & -- & \cellcolor{gray!12}\checkmark
& 29.4\rminus{5.1}
& 41.5\rminus{5.5}
& 42.6\rminus{18.3}
& 63.9\rminus{0.8}
& 29.7\rminus{29.8}
& \cellcolor{gray!12}78.4\gplus{20.5}
& 47.6\rminus{6.5} \\

\midrule
\rowcolor{oursbg}
\#8 & \checkmark & \checkmark & \checkmark
& \checkmark & \checkmark & \checkmark
& \textbf{66.8} & \textbf{72.8} & \textbf{72.9}
& \textbf{77.6} & \textbf{81.8} & \textbf{81.5}
& \textbf{75.6} \\

\#9 & \cellcolor{gray!12}-- & \checkmark & \checkmark
& \checkmark & \checkmark & \checkmark
& \cellcolor{gray!12}30.2\rminus{36.6}
& 74.4\gplus{1.6}
& 72.8\rminus{0.1}
& 77.9\gplus{0.3}
& 82.6\gplus{0.8}
& 80.7\rminus{0.8}
& 69.8\rminus{5.8} \\

\#10 & \checkmark & \cellcolor{gray!12}-- & \checkmark
& \checkmark & \checkmark & \checkmark
& 65.7\rminus{1.1}
& \cellcolor{gray!12}45.3\rminus{27.5}
& 73.1\gplus{0.2}
& 77.4\rminus{0.2}
& 82.6\gplus{0.8}
& 80.3\rminus{1.2}
& 70.7\rminus{4.9} \\

\#11 & \checkmark & \checkmark & \cellcolor{gray!12}--
& \checkmark & \checkmark & \checkmark
& 65.7\rminus{1.1}
& 75.1\gplus{2.3}
& \cellcolor{gray!12}60.8\rminus{12.1}
& 77.3\rminus{0.3}
& 82.4\gplus{0.6}
& 79.4\rminus{2.1}
& 73.5\rminus{2.1} \\

\#12 & \checkmark & \checkmark & \checkmark
& \cellcolor{gray!12}-- & \checkmark & \checkmark
& 65.5\rminus{1.3}
& 75.0\gplus{2.2}
& 72.4\rminus{0.5}
& \cellcolor{gray!12}67.5\rminus{10.1}
& 83.5\gplus{1.7}
& 81.1\rminus{0.4}
& 74.2\rminus{1.4} \\

\#13 & \checkmark & \checkmark & \checkmark
& \checkmark & \cellcolor{gray!12}-- & \checkmark
& 65.4\rminus{1.4}
& 75.0\gplus{2.2}
& 72.9\,{\scriptsize\textcolor{gray}{(0.0)}}
& 77.8\gplus{0.2}
& \cellcolor{gray!12}60.7\rminus{21.1}
& 80.4\rminus{1.1}
& 72.0\rminus{3.6} \\

\#14 & \checkmark & \checkmark & \checkmark
& \checkmark & \checkmark & \cellcolor{gray!12}--
& 64.8\rminus{2.0}
& 75.3\gplus{2.5}
& 72.5\rminus{0.4}
& 77.7\gplus{0.1}
& 83.0\gplus{1.2}
& \cellcolor{gray!12}60.8\rminus{20.7}
& 72.4\rminus{3.2} \\

\bottomrule
\end{tabular}%
}
\vskip -0.10in
\end{table*}

\begin{table}[t]
\centering
\caption{Ablation study of adopted and explored optimization strategies. Relative changes are shown alongside each metric, with green indicating improvements and red indicating degradations.}
\vskip -0.10in
\label{tab:3}
\setlength{\tabcolsep}{3pt}
\renewcommand{\arraystretch}{1.1}
\resizebox{\columnwidth}{!}{%
\begin{tabular}{@{}lcccc@{}}
\toprule
\multirow{2}{*}{Variant}
& \multicolumn{4}{c}{Avg.} \\
\cmidrule(lr){2-5}
& Acc $\uparrow$
& F1 $\uparrow$
& ECE $\downarrow$
& Brier $\downarrow$ \\
\midrule

\rowcolor{oursbg}
\textbf{UEmo-D1}
& \textbf{75.6}
& \textbf{69.5}
& \textbf{4.7}
& \textbf{33.5} \\

\quad $-$ Bidirectional Attn.
& 74.9\rminus{0.7}
& 69.1\rminus{0.4}
& 4.8\rplus{0.1}
& 33.6\rplus{0.1} \\

\quad $-$ Option-wise Reg.
& 75.2\rminus{0.4}
& 69.1\rminus{0.4}
& 4.7\dzero
& 33.7\rplus{0.2} \\

\quad $-$ Temperature Scaling
& 75.6\dzero
& 69.5\dzero
& 5.0\rplus{0.3}
& 33.5\dzero \\

\midrule

\quad $+$ 1 Extra Epoch
& 74.8\rminus{0.8}
& 69.3\rminus{0.2}
& 10.5\rplus{5.8}
& 36.9\rplus{3.4} \\

\quad $+$ 2 Extra Epochs
& 74.6\rminus{1.0}
& 69.2\rminus{0.3}
& 14.2\rplus{9.5}
& 39.3\rplus{5.8} \\

\quad $+$ Brier Loss
& 75.6\dzero
& 69.4\rminus{0.1}
& 5.1\rplus{0.4}
& 33.6\rplus{0.1} \\

\quad $+$ R-Drop~\cite{liang2021rdrop}
& 75.4\rminus{0.2}
& 69.4\rminus{0.1}
& 5.0\rplus{0.3}
& 33.6\rplus{0.1} \\

\quad $+$ GRPO~\cite{shao2024deepseekmath}
& 75.7\gplus{0.1}
& 69.5\dzero
& 9.3\rplus{4.6}
& 40.0\rplus{6.5} \\

\quad $+$ RLCR~\cite{damani2026beyond}
& 75.3\rminus{0.3}
& 69.0\rminus{0.5}
& 7.6\rplus{2.9}
& 37.0\rplus{3.5} \\

\bottomrule
\end{tabular}%
}
\vskip -0.10in
\end{table}

\section{Experiments}
\subsection{Preparations}
Following the training procedure described above, we train UEmo-D1 for one epoch on globally shuffled data, using a learning rate of $10^{-4}$ with 500 warmup steps followed by cosine decay, a LoRA rank of 16, and a batch size of 64. Training takes approximately 20 hours on four NVIDIA RTX 4090 GPUs. Post-training temperature scaling uses 10 equal-width confidence bins, yielding $\tau^{*}=1.079$.

With UEmo-D1 in place, we conduct a comprehensive experimental evaluation against open-source MLLMs~\cite{xu2025qwen25omni,xu2025qwen3omni,cui2026minicpmo45}, proprietary models~\cite{alibaba2026qwen3vlplus,moonshot2026kimik26,google2026gemini35flash,openai2026gpt56luna}, and emotion-oriented MLLMs~\cite{lian2025affectgpt,cheng2024emotionllama,huang2026nanoemox,wu2026emobserver,huang2026oneemo}.
We consider two inference modes where supported: native reasoning mode (R), which allows intermediate reasoning before producing an answer, and decision mode (D), which selects an answer by comparing candidate-option probabilities at the first output-token position.
We evaluate both modes on UEmo-Data's test split, comprising 41 subsets derived from 4 FER datasets~\cite{lee2019context,jiang2020dfew,barsoum2016training,goodfellow2013challenges}, 3 AER datasets~\cite{cao2014cremad,livingstone2018ravdess,busso2008iemocap}, 7 VCEC datasets~\cite{wu2026mvei,machajdik2010affective,yang2023emoset,you2016building,panda2018contemplating}, 3 CMER datasets~\cite{zadeh2018multimodal,cao2014cremad,poria2019meld}, 3 ITSA datasets~\cite{jia2022beyond,niu2016sentiment,yu2019adapting}, and 4 MSU datasets~\cite{cai2019multimodal,zhang2022mintrec,castro2019sarcasm,hasan2019urfunny}.
Unless otherwise specified, all evaluation prompts feature diversified instruction phrasing and shuffled option orders, with identical configurations used across models.

We report four metrics: accuracy (Acc), macro-averaged F1 score (F1), expected calibration error (ECE), and Brier score. ECE measures the sample-weighted discrepancy between accuracy and confidence across confidence bins. The Brier score measures the mean squared distance between predicted probability distributions and one-hot ground-truth labels. Higher Acc and F1 and lower ECE and Brier scores indicate better performance.

\subsection{Main Comparisons}

\noindent
\textbf{Comparison of Recognition Accuracy.}
Table~\ref{tab:1} compares representative models in native reasoning mode and a decision mode analogous to UEmo-D1. Overall, open-source and proprietary MLLMs exhibit relatively modest performance differences, underscoring the value of affect-specific optimization beyond general-purpose capabilities. Another notable finding is that decision mode outperforms reasoning mode in most comparisons, even for emotion-oriented MLLMs explicitly optimized for reasoning. This provides quantitative evidence that explicit reasoning may contribute more to explainability than to recognition accuracy itself. The results further reveal generalization challenges across both tasks and task formulations. Strengths such as OneEmo on FER and EmObserver on VCEC do not consistently generalize across tasks, and Nano-EmoX's strong reported performance does not carry over to our evaluation with diversified instructions and shuffled options. Against this backdrop, UEmo-D1 consistently outperforms all competing models across the six tasks, achieving accuracy gains of 4.8--18.4\% over the strongest baseline in each task. Its average accuracy reaches 75.6\%, exceeding the runner-up, Gemini-3.5-Flash, by 13.3\%.

\Cref{fig:5} further compares leading MLLMs across all 41 fine-grained evaluation subsets. Except for Gemini-3.5-Flash, all models use decision mode for their stronger overall performance. UEmo-D1 leads on most subsets, with gains exceeding 15\% on 10 subsets. Of the 11 subsets where it trails the strongest competing model, five show gaps below 0.5\%. These results further substantiate UEmo-D1's broad advantages in recognition accuracy.

\noindent
\textbf{Comparison of Confidence Estimation.}
\Cref{tab:2} compares confidence estimation across models, offering a practical perspective on the quality of their predictive distributions when ground-truth emotion distributions are unavailable. For a consistent comparison, we derive option probabilities from first-token logits in decision mode rather than extracting answer-token probabilities after reasoning as explored in~\cite{wu2025emocaliber}, since the latter are not always readily accessible in practice. Under this protocol, UEmo-D1 achieves markedly better calibration and probabilistic accuracy, with an average ECE of 4.7, approximately one-third of the next-best result, and a Brier score of 33.5, only 60\% of the next-best score. Together, these findings suggest that UEmo-D1's strengths extend to a distributional perspective on emotion recognition, assigning confidence more reliably across emotional alternatives rather than merely selecting a single categorical label.

\begin{figure*}[t]
    \centering
    \includegraphics[width=1\linewidth]{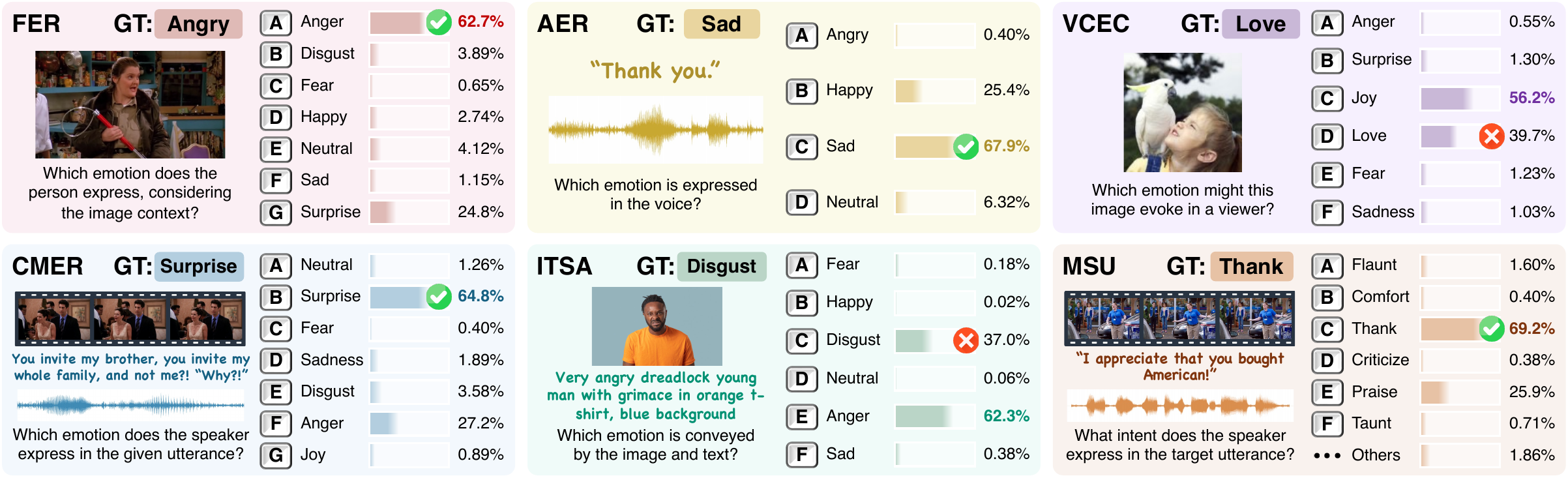}
    \vskip -0.10in
    \caption{Visualization of UEmo-D1's recognition results on representative cases from each of the six MER tasks. The multimodal content and instructions are shown on the left, where GT denotes the ground-truth label. The predicted probability distributions are shown on the right, with incorrect predictions for VCEC and ITSA and correct predictions for the remaining tasks.}
    \label{fig:6}
    \vskip -0.10in
\end{figure*}

\subsection{Further Analysis}

\noindent
\textbf{Influence of Optimization Strategies.} To examine the contributions of individual components,  \cref{tab:3} evaluates the removal of our three adopted modifications alongside alternative optimization strategies, each applied independently to UEmo-D1. Removing bidirectional attention or option-wise regularization slightly weakens recognition performance and probabilistic accuracy, while removing temperature scaling increases ECE without affecting classification accuracy. We attribute these modest effects partly to the large and diverse training collection, which may attenuate the gains from individual optimization refinements, as also suggested by prior observations on data scale~\cite{shi2024instruction}.

We also find no clear joint improvement from additional optimization strategies. Extra training epochs reduce accuracy and substantially worsen calibration, consistent with overfitting through repeated exposure to the same data. Regularization constraints, such as Brier loss and R-Drop, likewise offer no overall benefit. We further explore reinforcement learning, following its use in decision optimization~\cite{almeida2026jev,gao2026beyond}. GRPO improves accuracy by only 0.1\% while nearly doubling ECE, echoing its reported tendency toward overconfidence~\cite{bereket2025uncalibrated}. Explicit calibration rewards through RLCR partially alleviate this deterioration but still fall short of UEmo-D1. Together, these findings suggest that additional optimization complexity offers limited benefits over a simple supervised recipe for MER decision modeling under our training setup.

\noindent
\textbf{Correlation among MER Tasks.}
Beyond optimization strategies, we examine how different training tasks interact. The upper block of \cref{tab:4} compares models fine-tuned on individual tasks against the base model (\#1), revealing uneven cross-task transfer, with the largest gains generally concentrated on the task used for training. Interestingly, training on most other tasks improves ITSA and MSU, whereas the reverse is less consistent. This asymmetry suggests that these two tasks may require higher-level understanding and integration that benefit from improvements in more fundamental MER capabilities.

Conversely, the lower block examines the effects of excluding each task from UEmo-D1 (\#8). Every exclusion reduces average accuracy, highlighting the contribution of each task to overall performance. More specifically, all exclusions impair FER and MSU, whereas AER and ITSA improve whenever a task other than themselves is removed. These patterns do not simply mirror the single-task results, suggesting that transfer depends on the composition of the training mixture. Overall, these findings highlight both the value of joint training and the interplay between task complementarity and interference. While a deeper investigation could inform more effective data composition, the scope and space constraints of the present study leave this promising direction for future work.

\noindent
\textbf{Case Analysis.}
\Cref{fig:6} visualizes representative cases for an intuitive illustration of UEmo-D1's decision behavior. Even when predicting correctly, UEmo-D1 assigns appreciable probability to plausible alternatives, such as anger alongside surprise in CMER and praise alongside gratitude in MSU. Although the VCEC and ITSA predictions disagree with the ground-truth labels, their distributions capture plausible ambiguity between joy and love, and between anger and disgust, respectively. These examples illustrate how probabilistic decisions can reflect competing emotional interpretations beyond a single categorical label, even when the top-ranked prediction is counted as incorrect.
\section{Conclusion}
In this work, we introduce UEmo-D1, the first decision model for unified MER, combining efficient inference, superior performance, and enhanced calibration. To support systematic evaluation and joint optimization, we also curate UEmo-Data, a comprehensive data resource encompassing 41 evaluation subtasks and nearly one million training examples spanning six MER tasks. Beyond demonstrating UEmo-D1’s strengths across tasks, our systematic experiments shed light on how training strategies and task composition shape model performance. Together, we hope that UEmo-D1 will provide a practical, reproducible baseline for future research on unified MER and contribute to broader efforts toward human-level emotion recognition.

\section*{LLM Usage Statement}

During this study, we used MLLM-based agents (OpenAI Codex, GPT-6 Astra) to assist with data-resource integration and downloading, training-code execution, and data organization for quantitative analyses. During manuscript preparation, we used Codex to refine the language of our original writing and assist with preparing figures and tables. All AI-assisted outputs were carefully reviewed and refined by the authors before inclusion in the manuscript.

{
    \small
    \bibliographystyle{ieeenat_fullname}
    \bibliography{main}

\begin{thebibliography}{9}
\providecommand{\natexlab}[1]{#1}
\providecommand{\url}[1]{\texttt{#1}}
\expandafter\ifx\csname urlstyle\endcsname\relax
  \providecommand{\doi}[1]{doi: #1}\else
  \providecommand{\doi}{doi: \begingroup \urlstyle{rm}\Url}\fi

\bibitem[Achlioptas et~al.(2023)Achlioptas, Ovsjanikov, Guibas, and Tulyakov]{achlioptas2023affection}
Panos Achlioptas, Maks Ovsjanikov, Leonidas Guibas, and Sergey Tulyakov.
\newblock Affection: learning affective explanations for real-world visual data.
\newblock In \emph{CVPR}, pages 6641--6651, 2023.

\bibitem[Adigwe et~al.(2018)Adigwe, Tits, El~Haddad, Ostadabbas, and Dutoit]{adigwe2018emotional}
Adaeze Adigwe, No{\'e} Tits, Kevin El~Haddad, Sarah Ostadabbas, and Thierry Dutoit.
\newblock The emotional voices database: towards controlling the emotion dimension in voice generation systems.
\newblock \emph{arXiv preprint arXiv:1806.09514}, 2018.

\bibitem[Chen et~al.(2026)Chen, Guo, Chun, Wang, Perrault, and Elsner]{chen2026listen}
Jingyi Chen, Zhimeng Guo, Jiyun Chun, Pichao Wang, Andrew Perrault, and Micha Elsner.
\newblock Do audio {LLMs} really {LISTEN}, or just transcribe? measuring lexical vs. acoustic emotion cues reliance.
\newblock In \emph{EACL}, pages 5848--5877, 2026.

\bibitem[Han et~al.(2025)Han, Zhu, Xu, Song, and Yang]{han2025benchmarking}
Zhiyuan Han, Beier Zhu, Yanlong Xu, Peipei Song, and Xun Yang.
\newblock Benchmarking and bridging emotion conflicts for multimodal emotion reasoning.
\newblock In \emph{ACM MM}, pages 5528--5537, 2025.

\bibitem[Jin et~al.(2024)Jin, Jia, Wang, Li, Zhou, Zhou, Qin, and Wu]{jin2024speechcraft}
Zeyu Jin, Jia Jia, Qixin Wang, Kehan Li, Shuoyi Zhou, Songtao Zhou, Xiaoyu Qin, and Zhiyong Wu.
\newblock {SpeechCraft}: a fine-grained expressive speech dataset with natural language description.
\newblock In \emph{ACM MM}, pages 1255--1264, 2024.

\bibitem[Mertens et~al.(2024)Mertens, Yargholi, Op~de Beeck, Van~den Stock, and Vennekens]{mertens2024findingemo}
Laurent Mertens, Elahe' Yargholi, Hans Op~de Beeck, Jan Van~den Stock, and Joost Vennekens.
\newblock {FindingEmo}: an image dataset for emotion recognition in the wild.
\newblock In \emph{NeurIPS}, pages 4956--4996, 2024.

\bibitem[Pichora-Fuller and Dupuis(2020)]{pichorafuller2020tess}
M.~Kathleen Pichora-Fuller and Kate Dupuis.
\newblock {Toronto} emotional speech set ({TESS}).
\newblock Borealis, 2020.

\bibitem[Zhang et~al.(2018)Zhang, Luo, Loy, and Tang]{zhang2018facial}
Zhanpeng Zhang, Ping Luo, Chen~Change Loy, and Xiaoou Tang.
\newblock From facial expression recognition to interpersonal relation prediction.
\newblock \emph{IJCV}, 126\penalty0 (5):\penalty0 550--569, 2018.

\bibitem[Zhou et~al.(2022)Zhou, Sisman, Liu, and Li]{zhou2022emotional}
Kun Zhou, Berrak Sisman, Rui Liu, and Haizhou Li.
\newblock Emotional voice conversion: theory, databases and {ESD}.
\newblock \emph{Speech Communication}, 137:\penalty0 1--18, 2022.

\end{thebibliography}


\begin{thebibliography}{78}
\providecommand{\natexlab}[1]{#1}
\providecommand{\url}[1]{\texttt{#1}}
\expandafter\ifx\csname urlstyle\endcsname\relax
  \providecommand{\doi}[1]{doi: #1}\else
  \providecommand{\doi}{doi: \begingroup \urlstyle{rm}\Url}\fi

\bibitem[{Alibaba Cloud}(2026)]{alibaba2026qwen3vlplus}
{Alibaba Cloud}.
\newblock {Qwen3-VL-Plus}.
\newblock Alibaba Cloud Model Studio, 2026.
\newblock Online resource.

\bibitem[Almeida(2026)]{almeida2026jev}
Diogo Almeida.
\newblock Introducing {System One} models \& {Jev}.
\newblock TypeSafe AI Blog, 2026.
\newblock Online resource. Accessed November 16, 2026.

\bibitem[Bagher~Zadeh et~al.(2018)Bagher~Zadeh, Liang, Poria, Cambria, and Morency]{zadeh2018multimodal}
AmirAli Bagher~Zadeh, Paul~Pu Liang, Soujanya Poria, Erik Cambria, and Louis-Philippe Morency.
\newblock Multimodal language analysis in the wild: {CMU-MOSEI} dataset and interpretable dynamic fusion graph.
\newblock In \emph{ACL}, pages 2236--2246, 2018.

\bibitem[Barsoum et~al.(2016)Barsoum, Zhang, Canton~Ferrer, and Zhang]{barsoum2016training}
Emad Barsoum, Cha Zhang, Cristian Canton~Ferrer, and Zhengyou Zhang.
\newblock Training deep networks for facial expression recognition with crowd-sourced label distribution.
\newblock \emph{arXiv preprint arXiv:1608.01041}, 2016.

\bibitem[Bereket and Leskovec(2025)]{bereket2025uncalibrated}
Michael Bereket and Jure Leskovec.
\newblock Uncalibrated reasoning: {GRPO} induces overconfidence for stochastic outcomes.
\newblock \emph{arXiv preprint arXiv:2508.11800}, 2025.

\bibitem[Busso et~al.(2008)Busso, Bulut, Lee, Kazemzadeh, Mower, Kim, Chang, Lee, and Narayanan]{busso2008iemocap}
Carlos Busso, Murtaza Bulut, Chi-Chun Lee, Abe Kazemzadeh, Emily Mower, Samuel Kim, Jeannette~N. Chang, Sungbok Lee, and Shrikanth~S. Narayanan.
\newblock {IEMOCAP}: interactive emotional dyadic motion capture database.
\newblock \emph{Lang. Resour. Eval.}, 42\penalty0 (4):\penalty0 335--359, 2008.

\bibitem[Cai et~al.(2019)Cai, Cai, and Wan]{cai2019multimodal}
Yitao Cai, Huiyu Cai, and Xiaojun Wan.
\newblock Multi-modal sarcasm detection in {Twitter} with hierarchical fusion model.
\newblock In \emph{ACL}, pages 2506--2515, 2019.

\bibitem[Cao et~al.(2014)Cao, Cooper, Keutmann, Gur, Nenkova, and Verma]{cao2014cremad}
Houwei Cao, David~G. Cooper, Michael~K. Keutmann, Ruben~C. Gur, Ani Nenkova, and Ragini Verma.
\newblock {CREMA-D}: crowd-sourced emotional multimodal actors dataset.
\newblock \emph{IEEE Trans. Affective Comput.}, 5\penalty0 (4):\penalty0 377--390, 2014.

\bibitem[Castro et~al.(2019)Castro, Hazarika, P{\'e}rez-Rosas, Zimmermann, Mihalcea, and Poria]{castro2019sarcasm}
Santiago Castro, Devamanyu Hazarika, Ver{\'o}nica P{\'e}rez-Rosas, Roger Zimmermann, Rada Mihalcea, and Soujanya Poria.
\newblock Towards multimodal sarcasm detection (an obviously perfect paper).
\newblock In \emph{ACL}, pages 4619--4629, 2019.

\bibitem[Cheng et~al.(2024)Cheng, Cheng, He, Sun, Wang, Lin, Lian, Peng, and Hauptmann]{cheng2024emotionllama}
Zebang Cheng, Zhi-Qi Cheng, Jun-Yan He, Jingdong Sun, Kai Wang, Yuxiang Lin, Zheng Lian, Xiaojiang Peng, and Alexander~G. Hauptmann.
\newblock {Emotion-LLaMA}: multimodal emotion recognition and reasoning with instruction tuning.
\newblock In \emph{NeurIPS}, pages 110805--110853, 2024.

\bibitem[Cowen et~al.(2019)Cowen, Sauter, Tracy, and Keltner]{cowen2019mapping}
Alan Cowen, Disa Sauter, Jessica~L. Tracy, and Dacher Keltner.
\newblock Mapping the passions: toward a high-dimensional taxonomy of emotional experience and expression.
\newblock \emph{Psychol. Sci. Public Interest}, 20\penalty0 (1):\penalty0 69--90, 2019.

\bibitem[Cui et~al.(2026)Cui, Xu, Wang, Yu, Sun, Xu, Wang, et~al.]{cui2026minicpmo45}
Junbo Cui, Bokai Xu, Chongyi Wang, Tianyu Yu, Weiyue Sun, Yingjing Xu, Tianran Wang, et~al.
\newblock {MiniCPM-o 4.5}: towards real-time full-duplex omni-modal interaction.
\newblock \emph{arXiv preprint arXiv:2604.27393}, 2026.

\bibitem[Damani et~al.(2026)Damani, Puri, Slocum, Shenfeld, Choshen, Kim, and Andreas]{damani2026beyond}
Mehul Damani, Isha Puri, Stewart Slocum, Idan Shenfeld, Leshem Choshen, Yoon Kim, and Jacob Andreas.
\newblock Beyond binary rewards: training {LMs} to reason about their uncertainty.
\newblock In \emph{ICLR}, pages 59456--59493, 2026.

\bibitem[Devlin et~al.(2019)Devlin, Chang, Lee, and Toutanova]{devlin2019bert}
Jacob Devlin, Ming-Wei Chang, Kenton Lee, and Kristina Toutanova.
\newblock {BERT}: pre-training of deep bidirectional transformers for language understanding.
\newblock In \emph{NAACL-HLT}, pages 4171--4186, 2019.

\bibitem[Dibeklio{\u{g}}lu et~al.(2015)Dibeklio{\u{g}}lu, Hammal, Yang, and Cohn]{dibeklioglu2015multimodal}
Hamdi Dibeklio{\u{g}}lu, Zakia Hammal, Ying Yang, and Jeffrey~F. Cohn.
\newblock Multimodal detection of depression in clinical interviews.
\newblock In \emph{ICMI}, pages 307--310, 2015.

\bibitem[Gao et~al.(2026)Gao, Hu, Wang, Tao, and Wang]{gao2026beyond}
Zhimin Gao, Meibo Hu, Jiamian Wang, Zhiqiang Tao, and Pichao Wang.
\newblock Beyond outcome rewards: post-rationale distribution optimization for selective prediction in reasoning models.
\newblock \emph{arXiv preprint arXiv:2609.38850}, 2026.

\bibitem[Geuter et~al.(2026)Geuter, Cheng, and Alvarez-Melis]{geuter2026releasing}
Jonathan Geuter, Andrew Cheng, and David Alvarez-Melis.
\newblock Releasing the inner {Jev}: a simple recipe for open decision models.
\newblock \emph{arXiv preprint arXiv:2610.17205}, 2026.

\bibitem[Gideon et~al.(2021)Gideon, McInnis, and Mower~Provost]{gideon2021addog}
John Gideon, Melvin~G. McInnis, and Emily Mower~Provost.
\newblock Improving cross-corpus speech emotion recognition with adversarial discriminative domain generalization ({ADDoG}).
\newblock \emph{IEEE Trans. Affective Comput.}, 12\penalty0 (4):\penalty0 1055--1068, 2021.

\bibitem[Gneiting and Raftery(2007)]{gneiting2007strictly}
Tilmann Gneiting and Adrian~E. Raftery.
\newblock Strictly proper scoring rules, prediction, and estimation.
\newblock \emph{J. Am. Stat. Assoc.}, 102\penalty0 (477):\penalty0 359--378, 2007.

\bibitem[Goodfellow et~al.(2013)Goodfellow, Erhan, Carrier, Courville, Mirza, Hamner, Cukierski, Tang, Thaler, Lee, Zhou, Ramaiah, Feng, Li, Wang, Athanasakis, Shawe-Taylor, Milakov, Park, Ionescu, Popescu, Grozea, Bergstra, Xie, Romaszko, Xu, Chuang, and Bengio]{goodfellow2013challenges}
Ian~J. Goodfellow, Dumitru Erhan, Pierre~Luc Carrier, Aaron Courville, Mehdi Mirza, Ben Hamner, Will Cukierski, Yichuan Tang, David Thaler, Dong-Hyun Lee, Yingbo Zhou, Chetan Ramaiah, Fangxiang Feng, Ruifan Li, Xiaojie Wang, Dimitris Athanasakis, John Shawe-Taylor, Maxim Milakov, John Park, Radu Ionescu, Marius Popescu, Cristian Grozea, James Bergstra, Jingjing Xie, Lukasz Romaszko, Bing Xu, Zhang Chuang, and Yoshua Bengio.
\newblock Challenges in representation learning: a report on three machine learning contests.
\newblock \emph{arXiv preprint arXiv:1307.0414}, 2013.

\bibitem[{Google}(2026)]{google2026gemini35flash}
{Google}.
\newblock {Gemini 3.5 Flash}.
\newblock Google AI Studio, 2026.
\newblock Online resource.

\bibitem[Hasan et~al.(2019)Hasan, Rahman, Bagher~Zadeh, Zhong, Tanveer, Morency, and Hoque]{hasan2019urfunny}
Md~Kamrul Hasan, Wasifur Rahman, AmirAli Bagher~Zadeh, Jianyuan Zhong, Md~Iftekhar Tanveer, Louis-Philippe Morency, and Mohammed~(Ehsan) Hoque.
\newblock {UR-FUNNY}: a multimodal language dataset for understanding humor.
\newblock In \emph{EMNLP-IJCNLP}, pages 2046--2056, 2019.

\bibitem[Hu et~al.(2022)Hu, Shen, Wallis, Allen-Zhu, Li, Wang, Wang, and Chen]{hu2022lora}
Edward~J. Hu, Yelong Shen, Phillip Wallis, Zeyuan Allen-Zhu, Yuanzhi Li, Shean Wang, Lu Wang, and Weizhu Chen.
\newblock {LoRA}: low-rank adaptation of large language models.
\newblock In \emph{ICLR}, 2022.

\bibitem[Hu et~al.(2021)Hu, Liu, Zhao, and Jin]{hu2021mmgcn}
Jingwen Hu, Yuchen Liu, Jinming Zhao, and Qin Jin.
\newblock {MMGCN}: multimodal fusion via deep graph convolution network for emotion recognition in conversation.
\newblock In \emph{ACL-IJCNLP}, pages 5666--5675, 2021.

\bibitem[Huang et~al.(2026{\natexlab{a}})Huang, Lian, Zhang, Chen, Peng, and Wang]{huang2026oneemo}
Jiahao Huang, Zheng Lian, Jingyi Zhang, Zhide Chen, Xiaojiang Peng, and Shaonan Wang.
\newblock {OneEmo}: a unified multimodal reasoning model for emotion perception, understanding, and interaction.
\newblock \emph{arXiv preprint arXiv:2608.06013}, 2026{\natexlab{a}}.

\bibitem[Huang et~al.(2026{\natexlab{b}})Huang, Lin, Yang, Feng, Zhu, Yang, and Chen]{huang2026nanoemox}
Jiahao Huang, Fengyan Lin, Xuechao Yang, Chen Feng, Kexin Zhu, Xu Yang, and Zhide Chen.
\newblock {Nano-EmoX}: unifying multimodal emotional intelligence from perception to empathy.
\newblock In \emph{CVPR}, pages 22986--22997, 2026{\natexlab{b}}.

\bibitem[Jia et~al.(2022)Jia, He, Zhang, Uprety, Song, and Lioma]{jia2022beyond}
Ao Jia, Yu He, Yazhou Zhang, Sagar Uprety, Dawei Song, and Christina Lioma.
\newblock Beyond emotion: a multi-modal dataset for human desire understanding.
\newblock In \emph{NAACL-HLT}, pages 1512--1522, 2022.

\bibitem[Jiang et~al.(2020)Jiang, Zong, Zheng, Tang, Xia, Lu, and Liu]{jiang2020dfew}
Xingxun Jiang, Yuan Zong, Wenming Zheng, Chuangao Tang, Wanchuang Xia, Cheng Lu, and Jiateng Liu.
\newblock {DFEW}: a large-scale database for recognizing dynamic facial expressions in the wild.
\newblock In \emph{ACM MM}, pages 2881--2889, 2020.

\bibitem[Keltner and Haidt(1999)]{keltner1999social}
Dacher Keltner and Jonathan Haidt.
\newblock Social functions of emotions at four levels of analysis.
\newblock \emph{Cogn. Emot.}, 13\penalty0 (5):\penalty0 505--521, 1999.

\bibitem[Kumar et~al.(2021)Kumar, Khokher, Gupta, and Raman]{kumar2021hybrid}
Puneet Kumar, Vedanti Khokher, Yukti Gupta, and Balasubramanian Raman.
\newblock Hybrid fusion based approach for multimodal emotion recognition with insufficient labeled data.
\newblock In \emph{ICIP}, pages 314--318, 2021.

\bibitem[Lan et~al.(2025)Lan, Xue, Qi, Jiang, Lu, and Chua]{lan2025expllm}
Xing Lan, Jian Xue, Ji Qi, Dongmei Jiang, Ke Lu, and Tat-Seng Chua.
\newblock {ExpLLM}: towards chain of thought for facial expression recognition.
\newblock \emph{IEEE Trans. Multimedia}, 27:\penalty0 3069--3081, 2025.

\bibitem[Lee et~al.(2019)Lee, Kim, Kim, Park, and Sohn]{lee2019context}
Jiyoung Lee, Seungryong Kim, Sunok Kim, Jungin Park, and Kwanghoon Sohn.
\newblock Context-aware emotion recognition networks.
\newblock In \emph{ICCV}, pages 10143--10152, 2019.

\bibitem[Li et~al.(2024)Li, Yang, Zhao, Wang, Wang, Yang, Sun, Kou, Qian, and Zhang]{li2024corrkd}
Mingcheng Li, Dingkang Yang, Xiao Zhao, Shuaibing Wang, Yan Wang, Kun Yang, Mingyang Sun, Dongliang Kou, Ziyun Qian, and Lihua Zhang.
\newblock Correlation-decoupled knowledge distillation for multimodal sentiment analysis with incomplete modalities.
\newblock In \emph{CVPR}, pages 12458--12468, 2024.

\bibitem[Li and Wagle(2026)]{li2026llmasjev}
Yinheng Li and Justin Wagle.
\newblock {LLM-as-Jev}: {LLMs} are already {Jev}-style decision models---when and how to fine-tune them.
\newblock \emph{arXiv preprint arXiv:2610.02076}, 2026.

\bibitem[Li et~al.(2025)Li, Dao, Bao, Tan, Chen, Liu, and Kong]{li2024emola}
Yifan Li, Anh Dao, Wentao Bao, Zhen Tan, Tianlong Chen, Huan Liu, and Yu Kong.
\newblock Facial affective behavior analysis with instruction tuning.
\newblock In \emph{ECCV}, pages 165--186, 2025.
\newblock ECCV 2024 proceedings.

\bibitem[Li et~al.(2022)Li, Xu, Zhu, and Zhao]{li2022clmlf}
Zhen Li, Bing Xu, Conghui Zhu, and Tiejun Zhao.
\newblock {CLMLF}: a contrastive learning and multi-layer fusion method for multimodal sentiment detection.
\newblock In \emph{Findings of NAACL}, pages 2282--2294, 2022.

\bibitem[Lian et~al.(2023)Lian, Sun, Sun, Gu, Wen, Zhang, Chen, Xu, Xu, Chen, Chen, Liang, Li, Yi, Liu, and Tao]{lian2023emer}
Zheng Lian, Haiyang Sun, Licai Sun, Hao Gu, Zhuofan Wen, Siyuan Zhang, Shun Chen, Mingyu Xu, Ke Xu, Kang Chen, Lan Chen, Shan Liang, Ya Li, Jiangyan Yi, Bin Liu, and Jianhua Tao.
\newblock Explainable multimodal emotion recognition.
\newblock \emph{arXiv preprint arXiv:2306.15401}, 2023.

\bibitem[Lian et~al.(2025{\natexlab{a}})Lian, Chen, Chen, Sun, Sun, Ren, Cheng, Liu, Liu, Peng, Yi, and Tao]{lian2025affectgpt}
Zheng Lian, Haoyu Chen, Lan Chen, Haiyang Sun, Licai Sun, Yong Ren, Zebang Cheng, Bin Liu, Rui Liu, Xiaojiang Peng, Jiangyan Yi, and Jianhua Tao.
\newblock {AffectGPT}: a new dataset, model, and benchmark for emotion understanding with multimodal large language models.
\newblock In \emph{ICML}, pages 36993--37014, 2025{\natexlab{a}}.

\bibitem[Lian et~al.(2025{\natexlab{b}})Lian, Sun, Sun, Chen, Chen, Gu, Wen, Chen, Zhang, Yao, Liu, Liu, Liang, Li, Yi, and Tao]{lian2025ovmer}
Zheng Lian, Haiyang Sun, Licai Sun, Haoyu Chen, Lan Chen, Hao Gu, Zhuofan Wen, Shun Chen, Siyuan Zhang, Hailiang Yao, Bin Liu, Rui Liu, Shan Liang, Ya Li, Jiangyan Yi, and Jianhua Tao.
\newblock {OV-MER}: towards open-vocabulary multimodal emotion recognition.
\newblock In \emph{ICML}, pages 37015--37050, 2025{\natexlab{b}}.

\bibitem[Lian et~al.(2026)Lian, Zhang, Chen, Zhang, Liu, Wu, Chen, Li, Peng, He, and Tao]{lian2026affectgptrl}
Zheng Lian, Fan Zhang, Lan Chen, Yazhou Zhang, Rui Liu, Jinyang Wu, Haoyu Chen, Xiaobai Li, Xiaojiang Peng, Bin He, and Jianhua Tao.
\newblock {AffectGPT-RL}: revealing roles of reinforcement learning in open-vocabulary emotion recognition.
\newblock \emph{arXiv preprint arXiv:2605.06126}, 2026.

\bibitem[Liang et~al.(2021)Liang, Wu, Li, Wang, Meng, Qin, Chen, Zhang, and Liu]{liang2021rdrop}
Xiaobo Liang, Lijun Wu, Juntao Li, Yue Wang, Qi Meng, Tao Qin, Wei Chen, Min Zhang, and Tie-Yan Liu.
\newblock {R-Drop}: regularized dropout for neural networks.
\newblock In \emph{NeurIPS}, pages 10890--10905, 2021.

\bibitem[Livingstone and Russo(2018)]{livingstone2018ravdess}
Steven~R. Livingstone and Frank~A. Russo.
\newblock The {Ryerson} audio-visual database of emotional speech and song ({RAVDESS}): a dynamic, multimodal set of facial and vocal expressions in {North American English}.
\newblock \emph{PLOS ONE}, 13\penalty0 (5):\penalty0 e0196391, 2018.

\bibitem[Lopes et~al.(2003)Lopes, Salovey, and Straus]{lopes2003emotional}
Paulo~N Lopes, Peter Salovey, and Rebecca Straus.
\newblock Emotional intelligence, personality, and the perceived quality of social relationships.
\newblock \emph{Pers. Individ. Dif.}, 35\penalty0 (3):\penalty0 641--658, 2003.

\bibitem[Machajdik and Hanbury(2010)]{machajdik2010affective}
Jana Machajdik and Allan Hanbury.
\newblock Affective image classification using features inspired by psychology and art theory.
\newblock In \emph{ACM MM}, pages 83--92, 2010.

\bibitem[Mollahosseini et~al.(2019)Mollahosseini, Hasani, and Mahoor]{mollahosseini2019affectnet}
Ali Mollahosseini, Behzad Hasani, and Mohammad~H. Mahoor.
\newblock {AffectNet}: a database for facial expression, valence, and arousal computing in the wild.
\newblock \emph{IEEE Trans. Affective Comput.}, 10\penalty0 (1):\penalty0 18--31, 2019.

\bibitem[{Moonshot AI}(2026)]{moonshot2026kimik26}
{Moonshot AI}.
\newblock {Kimi K2.6}.
\newblock Hugging Face, 2026.
\newblock Online resource.

\bibitem[Niu et~al.(2016)Niu, Zhu, Pang, and El-Saddik]{niu2016sentiment}
Teng Niu, Shiai Zhu, Lei Pang, and Abdulmotaleb El-Saddik.
\newblock Sentiment analysis on multi-view social data.
\newblock In \emph{MMM}, pages 15--27, 2016.

\bibitem[Ong et~al.(2015)Ong, Zaki, and Goodman]{ong2015affective}
Desmond~C. Ong, Jamil Zaki, and Noah~D. Goodman.
\newblock Affective cognition: exploring lay theories of emotion.
\newblock \emph{Cognition}, 143:\penalty0 141--162, 2015.

\bibitem[{OpenAI}(2026)]{openai2026gpt56luna}
{OpenAI}.
\newblock {GPT-5.6 Luna}.
\newblock OpenAI API, 2026.
\newblock Online resource.

\bibitem[Pakdaman~Naeini et~al.(2015)Pakdaman~Naeini, Cooper, and Hauskrecht]{naeini2015obtaining}
Mahdi Pakdaman~Naeini, Gregory~F. Cooper, and Milos Hauskrecht.
\newblock Obtaining well calibrated probabilities using {Bayesian} binning.
\newblock In \emph{AAAI}, pages 2901--2907, 2015.

\bibitem[Panda et~al.(2018)Panda, Zhang, Li, Lee, Lu, and Roy-Chowdhury]{panda2018contemplating}
Rameswar Panda, Jianming Zhang, Haoxiang Li, Joon-Young Lee, Xin Lu, and Amit~K. Roy-Chowdhury.
\newblock Contemplating visual emotions: understanding and overcoming dataset bias.
\newblock In \emph{ECCV}, pages 579--595, 2018.

\bibitem[Poria et~al.(2017)Poria, Cambria, Bajpai, and Hussain]{poria2017review}
Soujanya Poria, Erik Cambria, Rajiv Bajpai, and Amir Hussain.
\newblock A review of affective computing: from unimodal analysis to multimodal fusion.
\newblock \emph{Inf. Fusion}, 37:\penalty0 98--125, 2017.

\bibitem[Poria et~al.(2019)Poria, Hazarika, Majumder, Naik, Cambria, and Mihalcea]{poria2019meld}
Soujanya Poria, Devamanyu Hazarika, Navonil Majumder, Gautam Naik, Erik Cambria, and Rada Mihalcea.
\newblock {MELD}: a multimodal multi-party dataset for emotion recognition in conversations.
\newblock In \emph{ACL}, pages 527--536, 2019.

\bibitem[Shao et~al.(2024)Shao, Wang, Zhu, Xu, Song, Bi, Zhang, Zhang, Li, Wu, and Guo]{shao2024deepseekmath}
Zhihong Shao, Peiyi Wang, Qihao Zhu, Runxin Xu, Junxiao Song, Xiao Bi, Haowei Zhang, Mingchuan Zhang, Y.~K. Li, Y. Wu, and Daya Guo.
\newblock {DeepSeekMath}: pushing the limits of mathematical reasoning in open language models.
\newblock \emph{arXiv preprint arXiv:2402.03300}, 2024.

\bibitem[Shi et~al.(2024)Shi, Yang, Wu, Aitchison, Yilmaz, and Lipani]{shi2024instruction}
Zhengyan Shi, Adam~X. Yang, Bin Wu, Laurence Aitchison, Emine Yilmaz, and Aldo Lipani.
\newblock Instruction tuning with loss over instructions.
\newblock In \emph{NeurIPS}, pages 69176--69205, 2024.

\bibitem[Skerry and Saxe(2014)]{skerry2014common}
Amy~E. Skerry and Rebecca Saxe.
\newblock A common neural code for perceived and inferred emotion.
\newblock \emph{J. Neurosci.}, 34\penalty0 (48):\penalty0 15997--16008, 2014.

\bibitem[Touvron et~al.(2023)Touvron, Martin, Stone, Albert, Almahairi, Babaei, Bashlykov, Batra, Bhargava, et~al.]{touvron2023llama2}
Hugo Touvron, Louis Martin, Kevin Stone, Peter Albert, Amjad Almahairi, Yasmine Babaei, Nikolay Bashlykov, Soumya Batra, Prajjwal Bhargava, et~al.
\newblock {Llama 2}: open foundation and fine-tuned chat models.
\newblock \emph{arXiv preprint arXiv:2307.09288}, 2023.

\bibitem[Tracy and Robins(2008)]{tracy2008automaticity}
Jessica~L. Tracy and Richard~W. Robins.
\newblock The automaticity of emotion recognition.
\newblock \emph{Emotion}, 8\penalty0 (1):\penalty0 81--95, 2008.

\bibitem[Tu et~al.(2026)Tu, Jing, Luo, Cambria, Li, and Xu]{tu2026satisfactory}
Geng Tu, Ran Jing, Xuan Luo, Erik Cambria, Wenjie Li, and Ruifeng Xu.
\newblock Is multimodal conversational emotion recognition satisfactory? exploring the gaps in performance, generalization, and confidence.
\newblock \emph{Pattern Recognit.}, 175:\penalty0 113087, 2026.

\bibitem[Van~Kleef(2009)]{vankleef2009emotions}
Gerben~A Van~Kleef.
\newblock How emotions regulate social life: the emotions as social information ({EASI}) model.
\newblock \emph{Curr. Dir. Psychol. Sci.}, 18\penalty0 (3):\penalty0 184--188, 2009.

\bibitem[Wang et~al.(2020)Wang, Peng, Yang, Lu, and Qiao]{wang2020scn}
Kai Wang, Xiaojiang Peng, Jianfei Yang, Shijian Lu, and Yu Qiao.
\newblock Suppressing uncertainties for large-scale facial expression recognition.
\newblock In \emph{CVPR}, pages 6897--6906, 2020.

\bibitem[Wu et~al.(2025)Wu, Yang, Zhao, Ma, and Zhou]{wu2025sentimental}
Daiqing Wu, Dongbao Yang, Sicheng Zhao, Can Ma, and Yu Zhou.
\newblock An empirical study on configuring in-context learning demonstrations for unleashing {MLLMs}' sentimental perception capability.
\newblock In \emph{ICML}, pages 67986--67999, 2025.

\bibitem[Wu et~al.(2026{\natexlab{a}})Wu, Yang, Ma, and Zhou]{wu2025emocaliber}
Daiqing Wu, Dongbao Yang, Can Ma, and Yu Zhou.
\newblock {EmoCaliber}: advancing reliable visual emotion comprehension via confidence verbalization and calibration.
\newblock \emph{Pattern Recognit.}, 179:\penalty0 113716, 2026{\natexlab{a}}.

\bibitem[Wu et~al.(2026{\natexlab{b}})Wu, Yang, Yao, Zhang, Ma, Zhou, and Zhao]{wu2026emobserver}
Daiqing Wu, Dongbao Yang, Jiashu Yao, Hongrui Zhang, Can Ma, Yu Zhou, and Sicheng Zhao.
\newblock {MVEI} \& {EmObserver}: empowering {MLLM}-oriented visual emotional intelligence via emotion statement judgement.
\newblock \emph{arXiv preprint arXiv:2607.21061}, 2026{\natexlab{b}}.

\bibitem[Wu et~al.(2026{\natexlab{c}})Wu, Yang, Zhao, Ma, and Zhou]{wu2026mvei}
Daiqing Wu, Dongbao Yang, Sicheng Zhao, Can Ma, and Yu Zhou.
\newblock Customizing visual emotion evaluation for {MLLMs}: an open-vocabulary, multifaceted, and scalable approach.
\newblock In \emph{ICLR}, pages 96462--96482, 2026{\natexlab{c}}.

\bibitem[Xie et~al.(2024)Xie, Peng, Tseng, Chen, Hsu, Shuai, and Cheng]{xie2024emovit}
Hongxia Xie, Chu-Jun Peng, Yu-Wen Tseng, Hung-Jen Chen, Chan-Feng Hsu, Hong-Han Shuai, and Wen-Huang Cheng.
\newblock {EmoVIT}: revolutionizing emotion insights with visual instruction tuning.
\newblock In \emph{CVPR}, pages 26596--26605, 2024.

\bibitem[Xu et~al.(2025{\natexlab{a}})Xu, Guo, He, Hu, He, Bai, Chen, Wang, Fan, Dang, Zhang, Wang, Chu, and Lin]{xu2025qwen25omni}
Jin Xu, Zhifang Guo, Jinzheng He, Hangrui Hu, Ting He, Shuai Bai, Keqin Chen, Jialin Wang, Yang Fan, Kai Dang, Bin Zhang, Xiong Wang, Yunfei Chu, and Junyang Lin.
\newblock {Qwen2.5-Omni} technical report.
\newblock \emph{arXiv preprint arXiv:2503.20215}, 2025{\natexlab{a}}.

\bibitem[Xu et~al.(2025{\natexlab{b}})Xu, Guo, Hu, Chu, Wang, He, Wang, et~al.]{xu2025qwen3omni}
Jin Xu, Zhifang Guo, Hangrui Hu, Yunfei Chu, Xiong Wang, Jinzheng He, Yuxuan Wang, et~al.
\newblock {Qwen3-Omni} technical report.
\newblock \emph{arXiv preprint arXiv:2509.17765}, 2025{\natexlab{b}}.

\bibitem[Xu and Mao(2017)]{xu2017multisentinet}
Nan Xu and Wenji Mao.
\newblock {MultiSentiNet}: a deep semantic network for multimodal sentiment analysis.
\newblock In \emph{CIKM}, pages 2399--2402, 2017.

\bibitem[Yang et~al.(2023)Yang, Huang, Ding, Lischinski, Cohen-Or, and Huang]{yang2023emoset}
Jingyuan Yang, Qirui Huang, Tingting Ding, Dani Lischinski, Danny Cohen-Or, and Hui Huang.
\newblock {EmoSet}: a large-scale visual emotion dataset with rich attributes.
\newblock In \emph{ICCV}, pages 20383--20394, 2023.

\bibitem[You et~al.(2016)You, Luo, Jin, and Yang]{you2016building}
Quanzeng You, Jiebo Luo, Hailin Jin, and Jianchao Yang.
\newblock Building a large scale dataset for image emotion recognition: the fine print and the benchmark.
\newblock In \emph{AAAI}, pages 308--314, 2016.

\bibitem[Yu and Jiang(2019)]{yu2019adapting}
Jianfei Yu and Jing Jiang.
\newblock Adapting {BERT} for target-oriented multimodal sentiment classification.
\newblock In \emph{IJCAI}, pages 5408--5414, 2019.

\bibitem[Zhang et~al.(2022)Zhang, Xu, Wang, Zhou, Zhao, and Teng]{zhang2022mintrec}
Hanlei Zhang, Hua Xu, Xin Wang, Qianrui Zhou, Shaojie Zhao, and Jiayan Teng.
\newblock {MIntRec}: a new dataset for multimodal intent recognition.
\newblock In \emph{ACM MM}, pages 1688--1697, 2022.

\bibitem[Zhang et~al.(2026)Zhang, Wu, Li, Liu, Wang, Zhou, and Zhao]{zhang2026merllm}
Hongrui Zhang, Daiqing Wu, Yangyang Li, Kuien Liu, Yuhui Wang, Yu Zhou, and Sicheng Zhao.
\newblock Multimodal emotion recognition with large language models.
\newblock In \emph{IJCAI}, pages 8088--8096, 2026.
\newblock Survey Track.

\bibitem[Zhao et~al.(2021{\natexlab{a}})Zhao, Li, and Jin]{zhao2021mmin}
Jinming Zhao, Ruichen Li, and Qin Jin.
\newblock Missing modality imagination network for emotion recognition with uncertain missing modalities.
\newblock In \emph{ACL-IJCNLP}, pages 2608--2618, 2021{\natexlab{a}}.

\bibitem[Zhao et~al.(2025)Zhao, Yang, Peng, Bai, Yao, Sun, Chen, Fu, Chen, Wei, and Bo]{zhao2025humanomni}
Jiaxing Zhao, Qize Yang, Yixing Peng, Detao Bai, Shimin Yao, Boyuan Sun, Xiang Chen, Shenghao Fu, Weixuan Chen, Xihan Wei, and Liefeng Bo.
\newblock {HumanOmni}: a large vision-speech language model for human-centric video understanding.
\newblock \emph{arXiv preprint arXiv:2501.15111}, 2025.

\bibitem[Zhao et~al.(2021{\natexlab{b}})Zhao, Jia, Yang, Ding, and Keutzer]{zhao2021emotion}
Sicheng Zhao, Guoli Jia, Jufeng Yang, Guiguang Ding, and Kurt Keutzer.
\newblock Emotion recognition from multiple modalities: fundamentals and methodologies.
\newblock \emph{IEEE Signal Process. Mag.}, 38\penalty0 (6):\penalty0 59--73, 2021{\natexlab{b}}.

\bibitem[Zhu et~al.(2026)Zhu, Zhang, Yu, Shao, Tan, and Nie]{zhu2026uniemo}
Yijie Zhu, Lingsen Zhang, Zitong Yu, Rui Shao, Tao Tan, and Liqiang Nie.
\newblock {UniEmo}: unifying emotional understanding and generation with learnable expert queries.
\newblock \emph{IEEE Trans. Image Process.}, 35:\penalty0 5165--5180, 2026.

\end{thebibliography}
}

\clearpage
\clearpage
\setcounter{page}{1}
\maketitlesupplementary

\setcounter{dbltopnumber}{3}
\renewcommand{\dbltopfraction}{0.95}
\renewcommand{\textfraction}{0.05}

\makeatletter
\newcommand{\datasetline}[1]{%
  \multirow{-#1}{*}[-2pt]{%
    \textcolor{gray!65}{%
      \rule{0.35pt}{%
        \dimexpr #1\ht\@arstrutbox
                +#1\dp\@arstrutbox-4pt\relax
      }%
    }%
  }%
}
\makeatother

\newcommand{\taskband}[1]{%
  \multicolumn{7}{@{}c@{}}{%
    \cellcolor{gray!15}\textbf{#1}%
  }%
}

\begin{table*}[t]
\centering
\caption{
Composition of the UEmo-Data test split across six MER tasks. I, V, A, and T denote image, video, audio, and text, respectively.
}
\vskip -0.10in
\label{tab:a1}
\setlength{\tabcolsep}{3pt}
\renewcommand{\arraystretch}{1.12}
\small
\resizebox{\textwidth}{!}{%
\begin{tabular}{
@{}
l@{\hspace{2pt}}c@{\hspace{4pt}}lrccr
@{}p{1.3em}@{}
l@{\hspace{2pt}}c@{\hspace{4pt}}lrccr
@{}
}
\toprule
\textbf{Source Dataset} &
& \textbf{Subtask}
& \textbf{\#Samples}
& \textbf{Input}
& \textbf{\#Classes}
& \textbf{Option Type}
&
& \textbf{Source Dataset} &
& \textbf{Subtask}
& \textbf{\#Samples}
& \textbf{Input}
& \textbf{\#Classes}
& \textbf{Option Type} \\
\midrule

\taskband{FER}
&& \taskband{CMER} \\

CAER-S~\cite{lee2019context} & \datasetline{1}
& CAER-S & 500 & I,T & 7 & Emotion
&& &
& MOSEI Anger & 500 & V,A,T & 2 & Emotion \\

DFEW~\cite{jiang2020dfew} & \datasetline{1}
& DFEW & 500 & V,T & 7 & Emotion
&& &
& MOSEI Disgust & 500 & V,A,T & 2 & Emotion \\

FERPlus~\cite{barsoum2016training} & \datasetline{1}
& FERPlus & 500 & I,T & 8 & Emotion
&& &
& MOSEI Fear & 500 & V,A,T & 2 & Emotion \\

FER2013~\cite{goodfellow2013challenges} & \datasetline{1}
& FER2013 & 500 & I,T & 7 & Emotion
&& &
& MOSEI Happiness & 500 & V,A,T & 2 & Emotion \\

\taskband{AER}
&& &
& MOSEI Sadness & 500 & V,A,T & 2 & Emotion \\

CREMA-D~\cite{cao2014cremad} & \datasetline{1}
& CREMA-D & 500 & A,T & 6 & Emotion
&& &
& MOSEI Surprise & 500 & V,A,T & 2 & Emotion \\

&
& RAVDESS Song & 176 & A,T & 6 & Emotion
&& \multirow{-7}{*}{CMU-MOSEI~\cite{zadeh2018multimodal}}
& \datasetline{7}
& MOSEI Sentiment & 500 & V,A,T & 3 & Sentiment \\

\multirow{-2}{*}{RAVDESS~\cite{livingstone2018ravdess}}
& \datasetline{2}
& RAVDESS Speech & 240 & A,T & 8 & Emotion
&& &
& MELD Emotion & 500 & V,A,T & 7 & Emotion \\

IEMOCAP~\cite{busso2008iemocap} & \datasetline{1}
& IEMOCAP & 500 & A,T & 4 & Emotion
&& \multirow{-2}{*}{MELD~\cite{poria2019meld}}
& \datasetline{2}
& MELD Sentiment & 500 & V,A,T & 3 & Sentiment \\

\taskband{VCEC}
&& CREMA-D~\cite{cao2014cremad} & \datasetline{1}
& CREMA-D & 500 & V,A,T & 6 & Emotion \\

&
& MVEI-EI & 500 & I,T & 2 & Judgment
&& \taskband{ITSA} \\

&
& MVEI-PS & 500 & I,T & 2 & Judgment
&& &
& MSED Desire & 500 & I,T & 7 & Desire \\

&
& MVEI-SC & 500 & I,T & 2 & Judgment
&& &
& MSED Emotion & 500 & I,T & 6 & Emotion \\

\multirow{-4}{*}{MVEI~\cite{wu2026mvei}}
& \datasetline{4}
& MVEI-SP & 399 & I,T & 2 & Judgment
&& \multirow{-3}{*}{MSED~\cite{jia2022beyond}}
& \datasetline{3}
& MSED Sentiment & 500 & I,T & 3 & Sentiment \\

Abstract~\cite{machajdik2010affective} & \datasetline{1}
& Abstract-8 & 229 & I,T & 8 & Emotion
&& MVSA-Multiple~\cite{niu2016sentiment} & \datasetline{1}
& MVSA-Multiple & 500 & I,T & 3 & Sentiment \\

ArtPhoto~\cite{machajdik2010affective} & \datasetline{1}
& ArtPhoto-8 & 500 & I,T & 8 & Emotion
&& Twitter-2017~\cite{yu2019adapting} & \datasetline{1}
& Twitter-2017 & 500 & I,T & 3 & Sentiment \\

EmoSet~\cite{yang2023emoset} & \datasetline{1}
& EmoSet-8 & 500 & I,T & 8 & Emotion
&& \taskband{MSU} \\

&
& FI-2 & 500 & I,T & 2 & Sentiment
&& HFM~\cite{cai2019multimodal} & \datasetline{1}
& HFM & 500 & I,T & 2 & Sarcasm \\

\multirow{-2}{*}{FI~\cite{you2016building}}
& \datasetline{2}
& FI-8 & 500 & I,T & 8 & Emotion
&& &
& MIntRec Coarse & 445 & V,A,T & 2 & Intention \\

&
& WebEmo-2 & 500 & I,T & 2 & Sentiment
&& \multirow{-2}{*}{MIntRec~\cite{zhang2022mintrec}}
& \datasetline{2}
& MIntRec Fine & 445 & V,A,T & 20 & Intention \\

&
& WebEmo-25 & 500 & I,T & 25 & Emotion
&& MUStARD~\cite{castro2019sarcasm} & \datasetline{1}
& MUStARD & 138 & V,A,T & 2 & Sarcasm \\

\multirow{-3}{*}{WebEmo~\cite{panda2018contemplating}}
& \datasetline{3}
& WebEmo-7 & 500 & I,T & 7 & Emotion
&& UR-FUNNY~\cite{hasan2019urfunny} & \datasetline{1}
& UR-FUNNY & 500 & V,A,T & 2 & Humor \\

UnbiasedEmo~\cite{panda2018contemplating} & \datasetline{1} 
& Unbiased Emo-6 & 500 & I,T & 6 & Emotion
&& \textbf{Overall} & \datasetline{1}
& \textbf{41 subsets}
& \textbf{19,072}
& I,V,A,T
& 2--25
& Mixed \\
\bottomrule
\end{tabular}%
}
\end{table*}

\newcommand{\traintaskband}[1]{%
  \multicolumn{6}{@{}c@{}}{%
    \cellcolor{gray!15}\textbf{#1}%
  }%
}

\begin{table*}[t]
\centering
\caption{
Composition of the UEmo-Data training split.
\#Instr. counts distinct instruction formulations within each data source.
}
\vskip -0.10in
\label{tab:a2}
\setlength{\tabcolsep}{3pt}
\renewcommand{\arraystretch}{1.12}
\small
\resizebox{\textwidth}{!}{%
\begin{tabular}{
@{}
>{\raggedright\arraybackslash}p{2.80cm}
>{\raggedleft\arraybackslash}p{1.35cm}
>{\centering\arraybackslash}p{1.05cm}
>{\centering\arraybackslash}p{1.25cm}
>{\centering\arraybackslash}p{1.35cm}
>{\raggedleft\arraybackslash}p{2.55cm}
@{}p{1.3em}@{}
>{\raggedright\arraybackslash}p{2.80cm}
>{\raggedleft\arraybackslash}p{1.35cm}
>{\centering\arraybackslash}p{1.05cm}
>{\centering\arraybackslash}p{1.25cm}
>{\centering\arraybackslash}p{1.35cm}
>{\raggedleft\arraybackslash}p{2.55cm}
@{}
}
\toprule
\textbf{Source Dataset}
& \textbf{\#Samples}
& \textbf{Input}
& \textbf{\#Classes}
& \textbf{\#Instr.}
& \textbf{Option Type}
&
& \textbf{Source Dataset}
& \textbf{\#Samples}
& \textbf{Input}
& \textbf{\#Classes}
& \textbf{\#Instr.}
& \textbf{Option Type} \\
\midrule

\traintaskband{FER}
&& \traintaskband{VCEC} \\

CAER-S~\cite{lee2019context}
& 49,007 & I,T & 7 & 200 & Emotion
&& INSETS-462K~\cite{wu2026emobserver}
& 232,213 & I,T & 2 & 200 & Judgment \\

DFEW~\cite{jiang2020dfew}
& 9,356 & V,T & 7 & 200 & Emotion
&& EmoSet~\cite{yang2023emoset}
& 94,474 & I,T & 8 & 200 & Emotion \\

FERPlus~\cite{barsoum2016training}
& 24,723 & I,T & 8 & 200 & Emotion
&& FI~\cite{you2016building}
& 19,637 & I,T & 2, 8 & 400 & Emotion, Sentiment \\

FER2013~\cite{goodfellow2013challenges}
& 28,322 & I,T & 7 & 200 & Emotion
&& WebEmo~\cite{panda2018contemplating}
& 226,658 & I,T & 2, 7, 25 & 400 & Emotion, Sentiment \\

ExpW~\citeapp{zhang2018facial}
& 88,600 & I,T & 7 & 200 & Emotion
&& Affection~\citeapp{achlioptas2023affection}
& 4,910 & I,T & 8 & 24 & Emotion \\

\traintaskband{AER}
&& FindingEmo~\citeapp{mertens2024findingemo}
& 13,093 & I,T & 24 & 24 & Emotion \\

CREMA-D~\cite{cao2014cremad}
& 4,672 & A,T & 6 & 200 & Emotion
&& \traintaskband{ITSA} \\

RAVDESS~\cite{livingstone2018ravdess}
& 1,540 & A,T & 6, 8 & 400 & Emotion
&& MSED~\cite{jia2022beyond}
& 18,381 & I,T & 3, 6, 7 & 600 & Desire, Emo., Senti. \\

IEMOCAP~\cite{busso2008iemocap}
& 3,128 & A,T & 4 & 200 & Emotion
&& \mbox{MVSA-Multiple~\cite{niu2016sentiment}}
& 12,958 & I,T & 3 & 200 & Sentiment \\

TESS~\citeapp{pichorafuller2020tess}
& 2,600 & A,T & 7 & 200 & Emotion
&& Twitter-2017~\cite{yu2019adapting}
& 3,545 & I,T & 3 & 200 & Sentiment \\

SpeechCraft~\citeapp{jin2024speechcraft}
& 1,000 & A,T & 7 & 197 & Emotion
&& MVSA-Single~\cite{niu2016sentiment}
& 3,604 & I,T & 3 & 200 & Sentiment \\

EmoV-DB~\citeapp{adigwe2018emotional}
& 6,893 & A,T & 5 & 200 & Emotion, Style
&& Twitter-2015~\cite{yu2019adapting}
& 3,174 & I,T & 3 & 200 & Sentiment \\

LISTEN~\citeapp{chen2026listen}
& 6,626 & A,T & 6, 10, 16 & 24 & Emotion
&& \traintaskband{MSU} \\

ESD~\citeapp{zhou2022emotional}
& 23,650 & A,T & 5 & 24 & Emotion
&& HFM~\cite{cai2019multimodal}
& 19,695 & I,T & 2 & 200 & Sarcasm \\

\traintaskband{CMER}
&& MIntRec~\cite{zhang2022mintrec}
& 2,668 & V,A,T & 2, 20 & 400 & Intention \\

CMU-MOSEI~\cite{zadeh2018multimodal}
& 40,197 & V,A,T & 2, 3 & 1,400 & Emotion, Sentiment
&& MUStARD~\cite{castro2019sarcasm}
& 499 & V,A,T & 2 & 198 & Sarcasm \\

MELD~\cite{poria2019meld}
& 19,026 & V,A,T & 3, 7 & 400 & Emotion, Sentiment
&& UR-FUNNY~\cite{hasan2019urfunny}
& 7,612 & V,A,T & 2 & 200 & Humor \\

CREMA-D~\cite{cao2014cremad}
& 4,809 & V,A,T & 6 & 200 & Emotion
&& \textbf{Overall}
& \textbf{977,270} & I,V,A,T & 2--25
& \textbf{8,091} & Mixed \\

\bottomrule
\end{tabular}%
}
\vskip -0.05in
\end{table*}

\section{Details of UEmo-Data}
\subsection{Data Composition}

\Cref{tab:a1} summarizes the test split, which comprises 19,072 samples across 41 subtasks, with 500 samples retained per subtask wherever sufficient data are available. To accommodate their differing task requirements, these subtasks use four input configurations: image--text (I, T), video--text (V, T), audio--text (A, T), and video--audio--text (V, A, T). Beyond input diversity, the candidate label spaces range from 2 to 25 classes, allowing evaluation at different levels of classification granularity. Most subtasks involve emotion or sentiment categories, while others assess complementary aspects of affective understanding. Specifically, MVEI evaluates the validity of emotion-related statements, whereas subsets of MSU and MSED Desire concern communicative intentions, social meanings, and desires. Together, these variations provide a comprehensive assessment across input configurations, label spaces, and recognition targets.

To support this breadth of evaluation, the training split draws on a wider collection of data resources, as detailed in \cref{tab:a2}. It supplements the task coverage of the test split with additional label inventories and diverse instruction formulations, exposing the model to varied expressions of related recognition requirements. The resulting 977,270 training samples provide rich supervision for learning heterogeneous MER tasks within a unified framework.

\begin{table*}[t]
\centering
\caption{
Inference latency and output length of MLLMs.
Time denotes mean recorded latency in seconds, measured on two NVIDIA RTX 4080 GPUs for locally deployed models.
Words measure output length, excluding internal reasoning of proprietary models.
R and D denote native reasoning and decision modes, respectively.
Avg. is computed over available evaluated samples.
}
\label{tab:a3}
\vskip -0.10in
\setlength{\tabcolsep}{3pt}
\renewcommand{\arraystretch}{1.10}
\small
\resizebox{0.88\textwidth}{!}{%
\begin{tabular}{@{}lc*{14}{r}@{}}
\toprule
\multirow{2}{*}{\textbf{Model}}
& \multirow{2}{*}{\textbf{Mode}}
& \multicolumn{2}{c}{FER}
& \multicolumn{2}{c}{AER}
& \multicolumn{2}{c}{VCEC}
& \multicolumn{2}{c}{CMER}
& \multicolumn{2}{c}{ITSA}
& \multicolumn{2}{c}{MSU}
& \multicolumn{2}{c}{Avg.} \\
\cmidrule(lr){3-4}
\cmidrule(lr){5-6}
\cmidrule(lr){7-8}
\cmidrule(lr){9-10}
\cmidrule(lr){11-12}
\cmidrule(lr){13-14}
\cmidrule(lr){15-16}
& & Time & Words & Time & Words & Time & Words
& Time & Words & Time & Words & Time & Words
& Time & Words \\
\midrule

\rowcolor{black!8}
\multicolumn{16}{c}{Open-Source MLLMs} \\

Qwen2.5-Omni~\cite{xu2025qwen25omni}
{\footnotesize\textcolor{gray}{7B}}
& R
& 1.211 & 21.3 & 1.317 & 23.6 & 1.791 & 32.8
& 3.484 & 11.2 & 1.705 & 30.6 & 2.899 & 12.8
& 2.226 & 22.9 \\

Qwen2.5-Omni~\cite{xu2025qwen25omni}
{\footnotesize\textcolor{gray}{7B}}
& D
& 0.166 & 1.0 & 0.081 & 1.0 & 0.183 & 1.0
& 2.898 & 1.0 & 0.168 & 1.0 & 2.107 & 1.0
& 1.066 & 1.0 \\

Qwen3-Omni~\cite{xu2025qwen3omni}
{\footnotesize\textcolor{gray}{30A3B}}
& R
& 0.953 & 1.0 & 0.911 & 1.0 & 1.067 & 1.0
& 2.555 & 1.0 & 1.076 & 1.0 & 2.076 & 1.4
& 1.528 & 1.0 \\

Qwen3-Omni~\cite{xu2025qwen3omni}
{\footnotesize\textcolor{gray}{30A3B}}
& D
& 0.852 & 1.0 & 0.810 & 1.0 & 0.931 & 1.0
& 2.373 & 1.0 & 0.946 & 1.0 & 1.774 & 1.0
& 1.368 & 1.0 \\

MiniCPM-o 4.5~\cite{cui2026minicpmo45}
{\footnotesize\textcolor{gray}{9B}}
& R
& 1.283 & 18.9 & 5.406 & 1.0 & 1.329 & 29.3
& 2.665 & 2.4 & 1.115 & 18.5 & 2.151 & 9.3
& 2.122 & 15.2 \\

MiniCPM-o 4.5~\cite{cui2026minicpmo45}
{\footnotesize\textcolor{gray}{9B}}
& D
& 0.163 & 1.0 & 0.088 & 1.0 & 0.060 & 1.0
& 1.318 & 1.0 & 0.060 & 1.0 & 1.056 & 1.0
& 0.501 & 1.0 \\

\rowcolor{black!8}
\multicolumn{16}{c}{Proprietary MLLMs} \\

Qwen3-VL-Plus~\cite{alibaba2026qwen3vlplus}
& R
& -- & 1.0 & -- & -- & -- & 1.0
& -- & 1.0 & -- & 1.0 & -- & 1.0
& -- & 1.0 \\

Kimi-K2.6~\cite{moonshot2026kimik26}
& R
& -- & 1.0 & -- & -- & -- & 1.0
& -- & 1.0 & -- & 1.0 & -- & 1.0
& -- & 1.0 \\

Gemini-3.5-Flash~\cite{google2026gemini35flash}
& R
& -- & 1.1 & -- & 1.0 & -- & 1.2
& -- & 1.0 & -- & 1.0 & -- & 1.0
& -- & 1.1 \\

GPT-5.6-Luna~\cite{openai2026gpt56luna}
& R
& -- & 1.0 & -- & -- & -- & 1.0
& -- & 1.0 & -- & 1.0 & -- & 1.0
& -- & 1.0 \\

\rowcolor{black!8}
\multicolumn{16}{c}{Emotion-Oriented MLLMs} \\

AffectGPT~\cite{lian2025affectgpt}
{\footnotesize\textcolor{gray}{7B}}
& R
& 0.925 & 1.3 & 1.133 & 1.5 & 1.232 & 2.3
& 1.940 & 1.1 & 0.850 & 2.0 & 1.721 & 2.6
& 1.378 & 1.8 \\

AffectGPT~\cite{lian2025affectgpt}
{\footnotesize\textcolor{gray}{7B}}
& D
& 0.145 & 1.0 & 0.113 & 1.0 & 0.094 & 1.0
& 1.102 & 1.0 & 0.095 & 1.0 & 0.643 & 1.0
& 0.414 & 1.0 \\

Emotion-LLaMA~\cite{cheng2024emotionllama}
{\footnotesize\textcolor{gray}{7B}}
& R
& 1.119 & 3.2 & 2.138 & 5.4 & 1.324 & 1.5
& 2.559 & 1.0 & 1.445 & 1.3 & 2.738 & 1.3
& 1.840 & 2.1 \\

Emotion-LLaMA~\cite{cheng2024emotionllama}
{\footnotesize\textcolor{gray}{7B}}
& D
& 0.500 & 1.0 & 0.668 & 1.0 & 0.511 & 1.0
& 2.303 & 1.0 & 0.512 & 1.0 & 2.252 & 1.0
& 1.114 & 1.0 \\

Nano-EmoX~\cite{huang2026nanoemox}
{\footnotesize\textcolor{gray}{1.5B}}
& R
& 5.041 & 10.9 & 4.242 & 27.3 & 3.364 & 14.9
& 3.588 & 10.4 & 2.558 & 33.6 & 4.624 & 21.2
& 3.670 & 18.1 \\

Nano-EmoX~\cite{huang2026nanoemox}
{\footnotesize\textcolor{gray}{1.5B}}
& D
& 0.221 & 1.0 & 0.178 & 1.0 & 0.155 & 1.0
& 1.892 & 1.0 & 0.154 & 1.0 & 1.087 & 1.0
& 0.701 & 1.0 \\

EmObserver~\cite{wu2026emobserver}
{\footnotesize\textcolor{gray}{8B}}
& R
& 21.610 & 548.4 & -- & -- & 25.881 & 643.2
& 20.903 & 453.5 & 39.350 & 1006.3 & 41.116 & 1007.7
& 26.505 & 643.1 \\

EmObserver~\cite{wu2026emobserver}
{\footnotesize\textcolor{gray}{8B}}
& D
& 0.150 & 1.0 & -- & --
& 0.134 & 1.0 & 3.667 & 1.0
& 0.116 & 1.0 & 2.482 & 1.0
& 1.405 & 1.0 \\

OneEmo~\cite{huang2026oneemo}
{\footnotesize\textcolor{gray}{4B}}
& R
& 5.555 & 86.1 & 5.740 & 83.5 & 5.927 & 89.8
& 6.888 & 80.4 & 5.802 & 88.3 & 5.960 & 80.3
& 6.134 & 85.4 \\

OneEmo~\cite{huang2026oneemo}
{\footnotesize\textcolor{gray}{4B}}
& D
& 0.310 & 1.0 & 0.209 & 1.0
& 0.289 & 1.0 & 4.062 & 1.0
& 0.279 & 1.0 & 2.706 & 1.0
& 1.497 & 1.0 \\

\rowcolor{blue!10}
\textbf{UEmo-D1 (Ours)}
{\footnotesize\textcolor{gray}{3B}}
& D
& 0.153 & 1.0 & 0.038 & 1.0 & 0.099 & 1.0
& 1.417 & 1.0 & 0.119 & 1.0 & 1.170 & 1.0
& 0.553 & 1.0 \\

\bottomrule
\end{tabular}%
}
\vskip -0.10in
\end{table*}

\subsection{Detailed Construction of UEmo-Data}

Building on the overview in \cref{sec:3}, we elaborate here on the procedures used to construct UEmo-Data.

\textbf{Label Mapping.}
In the source datasets, each sample consists of multimodal content and a ground-truth label under predefined task requirements and candidate label sets. To support probabilistic decisions at a single output-token position, we map each candidate option to an uppercase English letter. Specifically, we prefix each option with its assigned letter (e.g., ``A. Joy'') and incorporate this mapping into both the instruction format and the supervision target to ensure consistency between the presented options and the expected output.

\textbf{Deduplication.}
With the label mapping established, we perform content-based deduplication across modalities. Image screening uses decoded RGB content and perceptual similarity to identify identical or near-duplicate images. Audio screening similarly uses waveform identity and source metadata to detect repeated recordings and shared source utterances. Beyond redundancy, weakly labeled sources such as WebEmo and SpeechCraft require additional attention to annotation reliability and data composition. We exclude conflicting labels for identical images within the same WebEmo classification setting and limit the contribution of SpeechCraft's disproportionately large neutral-labeled collection. Specifically, its 352,261 records are reduced to 1,000 samples through deterministic round-robin selection across speakers and chapters, retaining 1,000 distinct speakers. Together, these adjustments reduce redundancy and source dominance while preserving diversity without enforcing uniform class distributions.

\textbf{Quality Control.}
Beyond deduplication, we address quality issues that could compromise evaluation reliability or introduce misleading supervision, focusing on media integrity and alignment as well as annotation consistency. We first check for missing or empty files, decoding failures, and mismatches between source records and their associated media, then correct temporal inconsistencies where necessary. For example, UR-FUNNY audio and video are reconstructed on a common timeline using their original timestamps to preserve synchronization despite differences in starting points or durations. Complementing these media-level checks, annotation screening follows criteria tailored to each source. In Affection, this requires the leading emotion to receive a strict majority of annotation votes and excludes images labeled ``something else.'' For INSETS-462K, we follow its established annotation-screening criteria by adopting the corresponding filtered collection while preserving the labels of retained records. The documented processing repairs alignment for approximately 7,600 audiovisual samples and filters out more than 160,000 source records, providing cleaner inputs and more reliable supervision for training and evaluation.

\textbf{Split Verification.}
The subsequent split assignment is relatively straightforward, but requires coordination across the collection. Many constituent datasets share multimodal sources, creating potential leakage when their independently defined partitions are combined. We retain datasets specifically intended for evaluation as test resources and, for others, select up to 500 samples per subtask from their original test partitions wherever available. To maximize separation between training and test content, we account for relationships at the speaker, dialogue, and source-video levels. RAVDESS uses a common actor-disjoint partition for speech and song, while CREMA-D similarly separates actors across splits. When incorporating LISTEN, we also exclude training records associated with held-out MELD dialogues or MOSEI source videos, preventing shared content from re-entering through another dataset. These measures help safeguard the validity of our evaluation.

\textbf{Instruction Formatting.}
Following split assignment, preliminary experiments revealed pronounced sensitivity to option order and limited cross-task transfer. We attribute these observations partly to fixed instruction formulations and option arrangements, consistent with the performance degradation of Nano-EmoX under our diversified evaluation in \cref{tab:1}. To mitigate these biases, we use LLMs to extensively reformulate task instructions, with cross-agent quality checks and human spot checks to preserve explicit and unambiguous recognition requirements. Complementing this linguistic diversity, candidate options are randomly permuted and their corresponding answer letters updated to reduce systematic associations between label identities and option positions. Together, these adjustments increase task-presentation diversity while reducing potential shortcuts. The resulting multimodal content \texttt{[MEDIA]}, instruction \texttt{[INST]}, and candidate options \texttt{[CHOICES]} are then assembled into a shared input template, which is applied uniformly across models evaluated in decision mode:
\begin{uemoprompt}
[MEDIA]
[INST]
Choose your answer from: A. [CHOICE 1]; B. [CHOICE 2]; C. [CHOICE 3]; ...
Answer with only the single option letter, without explanation.
\end{uemoprompt}

\section{Other Experiments}

\subsection{Comparison of Decision Efficiency}
Due to space constraints, the main paper provides only a brief account of inference efficiency. \Cref{tab:a3} complements \cref{fig:1} with task-level latency results and the average number of output words. These results confirm that UEmo-D1 offers competitive inference efficiency. We emphasize, however, that this efficiency largely follows from the decision-modeling paradigm and the use of a lightweight backbone, without dedicated efficiency optimizations. The same benefit is apparent across the compared models, which consistently run faster in decision mode than in their native mode.

It is also worth clarifying that ``reasoning mode'' refers here to a model's native response behavior, rather than an explicit requirement to reason. As a result, models trained to return concise predictions, such as AffectGPT and Emotion-LLaMA, can produce very short responses even in this mode. The roughly one-word outputs of proprietary models require a different interpretation: they represent only the visible answers, while any internal reasoning may be returned separately or remain hidden.

\begin{table*}[t]
\centering
\caption{
Accuracy comparison on the four FER subsets of UEmo-Data.
Open-source and emotion-oriented models are evaluated in decision
mode, while proprietary models use native inference.
Overall denotes the unweighted mean across the four subsets.
\textbf{Bold} and \underline{underline} indicate the best and
second-best results in each column, respectively.
}
\label{tab:a4}
\vskip -0.10in
\setlength{\tabcolsep}{6pt}
\renewcommand{\arraystretch}{1.12}
\small
\resizebox{0.66\textwidth}{!}{%
\begin{tabular}{
l
*{4}{>{\centering\arraybackslash}p{1.80cm}}
c
}
\toprule
\multirow{2}{*}{\textbf{Model}}
& \multicolumn{5}{c}{FER: Accuracy (\%) $\uparrow$} \\
\cmidrule(lr){2-6}
& CAER-S~\cite{lee2019context}
& DFEW~\cite{jiang2020dfew}
& FERPlus~\cite{barsoum2016training}
& FER2013~\cite{goodfellow2013challenges}
& Overall \\
\midrule

\rowcolor{black!8}
\multicolumn{6}{c}{Open-Source MLLMs} \\

Qwen2.5-Omni~\cite{xu2025qwen25omni}
{\footnotesize\textcolor{gray}{7B}}
& 29.4 & 44.0 & 42.0 & 37.6 & 38.2 \\

Qwen3-Omni~\cite{xu2025qwen3omni}
{\footnotesize\textcolor{gray}{30A3B}}
& 30.4 & 54.2 & 59.2 & 55.4 & 49.8 \\

MiniCPM-o 4.5~\cite{cui2026minicpmo45}
{\footnotesize\textcolor{gray}{9B}}
& 28.0 & 53.2 & 78.0 & 58.0 & 54.3 \\

\rowcolor{black!8}
\multicolumn{6}{c}{Proprietary MLLMs} \\

Qwen3-VL-Plus~\cite{alibaba2026qwen3vlplus}
& 31.2 & 56.4 & 74.8 & \underline{61.6} & 56.0 \\

Kimi-K2.6~\cite{moonshot2026kimik26}
& 29.8 & 52.6 & 70.4 & 55.6 & 52.1 \\

Gemini-3.5-Flash~\cite{google2026gemini35flash}
& \underline{31.9} & 50.0 & 67.6 & 56.6 & 51.5 \\

GPT-5.6-Luna~\cite{openai2026gpt56luna}
& 28.8 & 53.0 & 73.0 & 55.4 & 52.6 \\

\rowcolor{black!8}
\multicolumn{6}{c}{Emotion-Oriented MLLMs} \\

EmObserver~\cite{wu2026emobserver}
{\footnotesize\textcolor{gray}{8B}}
& 24.6 & 53.2 & 71.4 & 61.2 & 52.6 \\

OneEmo~\cite{huang2026oneemo}
{\footnotesize\textcolor{gray}{4B}}
& 27.2 & \underline{59.6} & \textbf{83.0} & 55.0
& \underline{56.2} \\

\rowcolor{blue!10}
\textbf{UEmo-D1 (Ours)}
{\footnotesize\textcolor{gray}{3B}}
& \textbf{51.6}
& \textbf{69.8}
& \underline{82.8}
& \textbf{63.0}
& \textbf{66.8} \\

\bottomrule
\end{tabular}}
\vskip -0.10in
\end{table*}

\begin{table*}[t]
\centering
\caption{
Accuracy comparison on the four AER subsets of UEmo-Data.
}
\label{tab:a5}
\vskip -0.10in
\setlength{\tabcolsep}{6pt}
\renewcommand{\arraystretch}{1.12}
\small
\resizebox{0.66\textwidth}{!}{%
\begin{tabular}{
l
*{4}{>{\centering\arraybackslash}p{1.80cm}}
c
}
\toprule
\multirow{3}{*}{\textbf{Model}}
& \multicolumn{5}{c}{AER: Accuracy (\%) $\uparrow$} \\
\cmidrule(lr){2-6}
& \multirow{2}{*}{CREMA-D~\cite{cao2014cremad}}
& \multicolumn{2}{c}{RAVDESS~\cite{livingstone2018ravdess}}
& \multirow{2}{*}{IEMOCAP~\cite{busso2008iemocap}}
& \multirow{2}{*}{Overall} \\
\cmidrule(lr){3-4}
& & Song & Speech & & \\
\midrule

\rowcolor{black!8}
\multicolumn{6}{c}{Open-Source MLLMs} \\

Qwen2.5-Omni~\cite{xu2025qwen25omni}
{\footnotesize\textcolor{gray}{7B}}
& 54.8
& \underline{51.7}
& \underline{46.3}
& 64.8
& \underline{54.4} \\

Qwen3-Omni~\cite{xu2025qwen3omni}
{\footnotesize\textcolor{gray}{30A3B}}
& 65.6 & 23.9 & 32.1 & 65.8 & 46.8 \\

MiniCPM-o 4.5~\cite{cui2026minicpmo45}
{\footnotesize\textcolor{gray}{9B}}
& 60.8 & 18.2 & 15.4 & 57.4 & 37.9 \\

\rowcolor{black!8}
\multicolumn{6}{c}{Proprietary MLLMs} \\

Gemini-3.5-Flash~\cite{google2026gemini35flash}
& \textbf{68.2}
& 35.8
& 39.2
& \underline{72.5}
& 53.9 \\

\rowcolor{black!8}
\multicolumn{6}{c}{Emotion-Oriented MLLMs} \\

OneEmo~\cite{huang2026oneemo}
{\footnotesize\textcolor{gray}{4B}}
& 54.4 & 18.2 & 6.7 & 30.6 & 27.5 \\

\rowcolor{blue!10}
\textbf{UEmo-D1 (Ours)}
{\footnotesize\textcolor{gray}{3B}}
& \underline{67.8}
& \textbf{70.5}
& \textbf{79.2}
& \textbf{73.6}
& \textbf{72.8} \\

\bottomrule
\end{tabular}}
\vskip -0.10in
\end{table*}

\begin{table*}[t]
\centering
\caption{
Accuracy comparison on the first group of VCEC subsets in UEmo-Data.
}
\label{tab:a6}
\vskip -0.10in
\setlength{\tabcolsep}{6pt}
\renewcommand{\arraystretch}{1.12}
\small
\resizebox{0.88\textwidth}{!}{%
\begin{tabular}{
l
*{7}{>{\centering\arraybackslash}p{1.60cm}}
}
\toprule
\multirow{3}{*}{\textbf{Model}}
& \multicolumn{7}{c}{VCEC: Accuracy (\%) $\uparrow$} \\
\cmidrule(lr){2-8}
& \multicolumn{4}{c}{MVEI~\cite{wu2026mvei}}
& \multicolumn{3}{c}{WebEmo~\cite{panda2018contemplating}} \\
\cmidrule(lr){2-5}
\cmidrule(lr){6-8}
& EI & PS & SC & SP
& 2-class & 7-class & 25-class \\
\midrule

\rowcolor{black!8}
\multicolumn{8}{c}{Open-Source MLLMs} \\

Qwen2.5-Omni~\cite{xu2025qwen25omni}
{\footnotesize\textcolor{gray}{7B}}
& 78.6 & 67.6 & 78.6 & 75.7
& 78.0 & 47.8 & 23.8 \\

Qwen3-Omni~\cite{xu2025qwen3omni}
{\footnotesize\textcolor{gray}{30A3B}}
& 78.8 & 73.8 & 82.0 & 76.4
& 73.2 & 48.0 & \underline{26.8} \\

MiniCPM-o 4.5~\cite{cui2026minicpmo45}
{\footnotesize\textcolor{gray}{9B}}
& 78.8 & 75.4 & 75.6 & 75.7
& 74.8 & 48.0 & 24.6 \\

\rowcolor{black!8}
\multicolumn{8}{c}{Proprietary MLLMs} \\

Qwen3-VL-Plus~\cite{alibaba2026qwen3vlplus}
& \underline{82.0} & 74.2 & 78.0 & \textbf{86.7}
& 75.6 & 47.4 & 24.0 \\

Kimi-K2.6~\cite{moonshot2026kimik26}
& 81.8 & 80.6 & 80.2 & 83.7
& 77.6 & 50.2 & 21.4 \\

Gemini-3.5-Flash~\cite{google2026gemini35flash}
& 75.3 & \underline{84.2} & 78.1 & 84.7
& \textbf{85.0} & 48.5 & 22.8 \\

GPT-5.6-Luna~\cite{openai2026gpt56luna}
& 75.4 & 72.2 & 74.8 & 77.2
& 71.4 & 44.4 & 22.6 \\

\rowcolor{black!8}
\multicolumn{8}{c}{Emotion-Oriented MLLMs} \\

EmObserver~\cite{wu2026emobserver}
{\footnotesize\textcolor{gray}{8B}}
& 78.4 & 79.8 & \underline{84.2} & 85.2
& 73.6 & \underline{53.0} & 26.6 \\

OneEmo~\cite{huang2026oneemo}
{\footnotesize\textcolor{gray}{4B}}
& 77.8 & 68.4 & 82.6 & 77.9
& 68.0 & 41.0 & 20.4 \\

\rowcolor{blue!10}
\textbf{UEmo-D1 (Ours)}
{\footnotesize\textcolor{gray}{3B}}
& \textbf{87.2}
& \textbf{84.6}
& \textbf{92.0}
& \underline{86.2}
& \underline{83.0}
& \textbf{62.0}
& \textbf{36.0} \\

\bottomrule
\end{tabular}%
}
\vskip -0.06in
\end{table*}

\begin{table*}[t]
\centering
\caption{
Accuracy comparison on the second group of VCEC subsets in UEmo-Data.
Overall includes results from \cref{tab:a6}.
}
\label{tab:a7}
\vskip -0.10in
\setlength{\tabcolsep}{7pt}
\renewcommand{\arraystretch}{1.12}
\small
\resizebox{0.88\textwidth}{!}{%
\begin{tabular}{
l
*{7}{>{\centering\arraybackslash}p{1.60cm}}
}
\toprule
\multirow{3}{*}{\textbf{Model}}
& \multicolumn{7}{c}{VCEC: Accuracy (\%) $\uparrow$} \\
\cmidrule(lr){2-8}
& \multirow{2}{*}{Abstract~\cite{machajdik2010affective}}
& \multirow{2}{*}{ArtPhoto~\cite{machajdik2010affective}}
& \multirow{2}{*}{EmoSet~\cite{yang2023emoset}}
& \multicolumn{2}{c}{FI~\cite{you2016building}}
& \multirow{2}{*}{UnbiasedEmo~\cite{panda2018contemplating}}
& \multirow{2}{*}{Overall} \\
\cmidrule(lr){5-6}
& & & & 2-class & 8-class & & \\
\midrule

\rowcolor{black!8}
\multicolumn{8}{c}{Open-Source MLLMs} \\

Qwen2.5-Omni~\cite{xu2025qwen25omni}
{\footnotesize\textcolor{gray}{7B}}
& 16.2 & 43.6 & 55.4 & 91.0
& 54.2 & 82.0 & 61.0 \\

Qwen3-Omni~\cite{xu2025qwen3omni}
{\footnotesize\textcolor{gray}{30A3B}}
& 29.3 & 45.2 & 60.6 & 92.4
& 59.2 & \textbf{88.8} & 64.2 \\

MiniCPM-o 4.5~\cite{cui2026minicpmo45}
{\footnotesize\textcolor{gray}{9B}}
& 24.9 & \underline{47.0} & 57.6 & 92.4
& 60.6 & 83.8 & 63.0 \\

\rowcolor{black!8}
\multicolumn{8}{c}{Proprietary MLLMs} \\

Qwen3-VL-Plus~\cite{alibaba2026qwen3vlplus}
& 19.2 & 46.0 & 55.2 & 93.2
& 56.0 & 83.6 & 63.2 \\

Kimi-K2.6~\cite{moonshot2026kimik26}
& 31.0 & 46.2 & 53.4 & 90.4
& 54.4 & 79.6 & 63.9 \\

Gemini-3.5-Flash~\cite{google2026gemini35flash}
& \underline{35.2} & \textbf{49.7} & 46.1
& \underline{94.5}
& 59.5 & 84.5 & 65.2 \\

GPT-5.6-Luna~\cite{openai2026gpt56luna}
& 18.3 & 43.8 & 54.8 & 93.2
& 54.4 & 83.0 & 60.4 \\

\rowcolor{black!8}
\multicolumn{8}{c}{Emotion-Oriented MLLMs} \\

EmObserver~\cite{wu2026emobserver}
{\footnotesize\textcolor{gray}{8B}}
& \textbf{43.2} & 46.0 & \underline{67.2} & 90.6
& \underline{72.0} & \underline{86.0}
& \underline{68.1} \\

OneEmo~\cite{huang2026oneemo}
{\footnotesize\textcolor{gray}{4B}}
& 11.4 & 38.2 & 41.6 & 88.4
& 47.8 & 80.6 & 57.2 \\

\rowcolor{blue!10}
\textbf{UEmo-D1 (Ours)}
{\footnotesize\textcolor{gray}{3B}}
& 34.1
& 41.4
& \textbf{80.8}
& \textbf{96.0}
& \textbf{78.2}
& \underline{86.0}
& \textbf{72.9} \\

\bottomrule
\end{tabular}%
}
\vskip -0.10in
\end{table*}

\begin{table*}[t]
\centering
\caption{
Accuracy comparison on the first group of CMER subsets in UEmo-Data.
}
\label{tab:a8}
\vskip -0.10in
\setlength{\tabcolsep}{6pt}
\renewcommand{\arraystretch}{1.12}
\small
\resizebox{0.88\textwidth}{!}{%
\begin{tabular}{
l
*{7}{>{\centering\arraybackslash}p{1.60cm}}
}
\toprule
\multirow{3}{*}{\textbf{Model}}
& \multicolumn{7}{c}{CMER: Accuracy (\%) $\uparrow$} \\
\cmidrule(lr){2-8}
& \multicolumn{7}{c}{CMU-MOSEI~\cite{zadeh2018multimodal}} \\
\cmidrule(lr){2-8}
& Anger & Disgust & Fear & Happiness
& Sadness & Surprise & Sentiment \\
\midrule

\rowcolor{black!8}
\multicolumn{8}{c}{Open-Source MLLMs} \\

Qwen2.5-Omni~\cite{xu2025qwen25omni}
{\footnotesize\textcolor{gray}{7B}}
& 80.4 & 87.0 & 89.2 & 66.8
& 70.8 & 86.8 & 65.8 \\

Qwen3-Omni~\cite{xu2025qwen3omni}
{\footnotesize\textcolor{gray}{30A3B}}
& \textbf{81.4} & \underline{87.6} & 87.4
& \underline{69.6}
& 70.6 & 90.4 & \underline{68.6} \\

MiniCPM-o 4.5~\cite{cui2026minicpmo45}
{\footnotesize\textcolor{gray}{9B}}
& 79.6 & 86.4 & 88.0 & 64.0
& 71.2 & 90.4 & 62.0 \\

\rowcolor{black!8}
\multicolumn{8}{c}{Proprietary MLLMs} \\

Qwen3-VL-Plus~\cite{alibaba2026qwen3vlplus}
& 79.6 & 84.2 & 87.6 & 68.0
& \textbf{72.4} & 70.8 & 64.8 \\

Kimi-K2.6~\cite{moonshot2026kimik26}
& 79.2 & 85.8 & 88.4 & 65.4
& 70.8 & 90.8 & 60.8 \\

Gemini-3.5-Flash~\cite{google2026gemini35flash}
& 76.5 & 80.6 & 82.5 & 61.1
& 64.6 & 85.5 & 55.1 \\

GPT-5.6-Luna~\cite{openai2026gpt56luna}
& 76.0 & 84.0 & 86.2 & 67.4
& 70.8 & 89.4 & 61.4 \\

\rowcolor{black!8}
\multicolumn{8}{c}{Emotion-Oriented MLLMs} \\

EmObserver~\cite{wu2026emobserver}
{\footnotesize\textcolor{gray}{8B}}
& 78.2 & 84.6 & \underline{89.6}
& \underline{69.6}
& 71.2 & 90.4 & 65.4 \\

OneEmo~\cite{huang2026oneemo}
{\footnotesize\textcolor{gray}{4B}}
& 77.8 & 84.4 & \textbf{89.8} & 56.4
& 70.4 & \textbf{91.8} & 60.6 \\

\rowcolor{blue!10}
\textbf{UEmo-D1 (Ours)}
{\footnotesize\textcolor{gray}{3B}}
& \underline{81.2}
& \textbf{88.0}
& 89.4
& \textbf{75.4}
& \underline{72.2}
& \underline{91.6}
& \textbf{73.0} \\

\bottomrule
\end{tabular}%
}
\vskip -0.10in
\end{table*}

\begin{table*}[t]
\centering
\caption{
Accuracy comparison on the second group of CMER subsets in UEmo-Data.
Overall includes results from \cref{tab:a8}.
}
\label{tab:a9}
\vskip -0.10in
\setlength{\tabcolsep}{6pt}
\renewcommand{\arraystretch}{1.12}
\small
\resizebox{0.56\textwidth}{!}{%
\begin{tabular}{
l
*{4}{>{\centering\arraybackslash}p{1.60cm}}
}
\toprule
\multirow{3}{*}{\textbf{Model}}
& \multicolumn{4}{c}{CMER: Accuracy (\%) $\uparrow$} \\
\cmidrule(lr){2-5}
& \multicolumn{2}{c}{MELD~\cite{poria2019meld}}
& \multirow{2}{*}{CREMA-D~\cite{cao2014cremad}}
& \multirow{2}{*}{Overall} \\
\cmidrule(lr){2-3}
& Emotion & Sentiment & & \\
\midrule

\rowcolor{black!8}
\multicolumn{5}{c}{Open-Source MLLMs} \\

Qwen2.5-Omni~\cite{xu2025qwen25omni}
{\footnotesize\textcolor{gray}{7B}}
& 50.2 & 51.2 & 54.8 & 70.3 \\

Qwen3-Omni~\cite{xu2025qwen3omni}
{\footnotesize\textcolor{gray}{30A3B}}
& \underline{55.4} & 49.4 & 57.0 & \underline{71.7} \\

MiniCPM-o 4.5~\cite{cui2026minicpmo45}
{\footnotesize\textcolor{gray}{9B}}
& 54.4 & 51.8 & 43.2 & 69.1 \\

\rowcolor{black!8}
\multicolumn{5}{c}{Proprietary MLLMs} \\

Qwen3-VL-Plus~\cite{alibaba2026qwen3vlplus}
& 44.4 & 55.4 & 58.2 & 68.5 \\

Kimi-K2.6~\cite{moonshot2026kimik26}
& 48.6 & \underline{63.8} & 61.6 & 71.5 \\

Gemini-3.5-Flash~\cite{google2026gemini35flash}
& 50.6 & 49.4 & \textbf{71.6} & 67.8 \\

GPT-5.6-Luna~\cite{openai2026gpt56luna}
& 35.4 & 46.2 & 61.2 & 67.8 \\

\rowcolor{black!8}
\multicolumn{5}{c}{Emotion-Oriented MLLMs} \\

EmObserver~\cite{wu2026emobserver}
{\footnotesize\textcolor{gray}{8B}}
& 34.2 & 38.4 & 43.4 & 66.5 \\

OneEmo~\cite{huang2026oneemo}
{\footnotesize\textcolor{gray}{4B}}
& 45.0 & 49.8 & 47.0 & 67.3 \\

\rowcolor{blue!10}
\textbf{UEmo-D1 (Ours)}
{\footnotesize\textcolor{gray}{3B}}
& \textbf{68.8}
& \textbf{74.6}
& \underline{62.2}
& \textbf{77.6} \\

\bottomrule
\end{tabular}%
}
\vskip -0.10in
\end{table*}

\begin{table*}[t]
\centering
\caption{
Accuracy comparison on the ITSA subsets of UEmo-Data.
}
\label{tab:a10}
\vskip -0.10in
\setlength{\tabcolsep}{6pt}
\renewcommand{\arraystretch}{1.12}
\small
\resizebox{0.76\textwidth}{!}{%
\begin{tabular}{
l
*{6}{>{\centering\arraybackslash}p{1.60cm}}
}
\toprule
\multirow{3}{*}{\textbf{Model}}
& \multicolumn{6}{c}{ITSA: Accuracy (\%) $\uparrow$} \\
\cmidrule(lr){2-7}
& \multicolumn{3}{c}{MSED~\cite{jia2022beyond}}
& MVSA~\cite{niu2016sentiment}
& Twitter~\cite{yu2019adapting}
& \multirow{2}{*}{Overall} \\
\cmidrule(lr){2-4}
\cmidrule(lr){5-5}
\cmidrule(lr){6-6}
& Desire & Emotion & Sentiment
& Multiple & 2017 & \\
\midrule

\rowcolor{black!8}
\multicolumn{7}{c}{Open-Source MLLMs} \\

Qwen2.5-Omni~\cite{xu2025qwen25omni}
{\footnotesize\textcolor{gray}{7B}}
& \underline{55.6} & 72.0 & 70.4
& \underline{70.4} & 52.0 & 64.1 \\

Qwen3-Omni~\cite{xu2025qwen3omni}
{\footnotesize\textcolor{gray}{30A3B}}
& 54.6 & 73.4 & 69.8
& 68.2 & 50.2 & 63.2 \\

MiniCPM-o 4.5~\cite{cui2026minicpmo45}
{\footnotesize\textcolor{gray}{9B}}
& 38.6 & 72.0 & 71.4
& 69.0 & 51.0 & 60.4 \\

\rowcolor{black!8}
\multicolumn{7}{c}{Proprietary MLLMs} \\

Qwen3-VL-Plus~\cite{alibaba2026qwen3vlplus}
& 51.0 & 72.8 & 70.2
& 69.2 & 54.4 & 63.5 \\

Kimi-K2.6~\cite{moonshot2026kimik26}
& 45.4 & \underline{76.6} & 71.6
& 69.2 & \underline{59.8} & \underline{64.5} \\

Gemini-3.5-Flash~\cite{google2026gemini35flash}
& 48.3 & 75.6 & \underline{72.6}
& 69.9 & 51.7 & 63.6 \\

GPT-5.6-Luna~\cite{openai2026gpt56luna}
& 51.0 & 71.6 & 70.8
& 70.2 & 53.6 & 63.4 \\

\rowcolor{black!8}
\multicolumn{7}{c}{Emotion-Oriented MLLMs} \\

EmObserver~\cite{wu2026emobserver}
{\footnotesize\textcolor{gray}{8B}}
& 49.6 & 69.0 & 70.0
& 69.6 & 46.8 & 61.0 \\

OneEmo~\cite{huang2026oneemo}
{\footnotesize\textcolor{gray}{4B}}
& 52.4 & 71.0 & 72.0
& 69.2 & 48.2 & 62.6 \\

\rowcolor{blue!10}
\textbf{UEmo-D1 (Ours)}
{\footnotesize\textcolor{gray}{3B}}
& \textbf{88.6}
& \textbf{88.0}
& \textbf{90.4}
& \textbf{72.2}
& \textbf{69.8}
& \textbf{81.8} \\

\bottomrule
\end{tabular}%
}
\vskip -0.10in
\end{table*}

\begin{table*}[t]
\centering
\caption{
Accuracy comparison on the MSU subsets of UEmo-Data.
}
\label{tab:a11}
\vskip -0.10in
\setlength{\tabcolsep}{5pt}
\renewcommand{\arraystretch}{1.12}
\small
\resizebox{0.76\textwidth}{!}{%
\begin{tabular}{
l
*{6}{>{\centering\arraybackslash}p{1.60cm}}
}
\toprule
\multirow{3}{*}{\textbf{Model}}
& \multicolumn{6}{c}{MSU: Accuracy (\%) $\uparrow$} \\
\cmidrule(lr){2-7}
& \multirow{2}{*}{HFM~\cite{cai2019multimodal}}
& \multicolumn{2}{c}{MIntRec~\cite{zhang2022mintrec}}
& \multirow{2}{*}{MUStARD~\cite{castro2019sarcasm}}
& \multirow{2}{*}{UR-FUNNY~\cite{hasan2019urfunny}}
& \multirow{2}{*}{Overall} \\
\cmidrule(lr){3-4}
& & Coarse & Fine & & & \\
\midrule

\rowcolor{black!8}
\multicolumn{7}{c}{Open-Source MLLMs} \\

Qwen2.5-Omni~\cite{xu2025qwen25omni}
{\footnotesize\textcolor{gray}{7B}}
& 74.4 & 74.6 & 63.4
& 63.0 & 60.2 & 67.1 \\

Qwen3-Omni~\cite{xu2025qwen3omni}
{\footnotesize\textcolor{gray}{30A3B}}
& 75.2 & \underline{77.1} & \underline{65.8}
& \underline{75.4} & 69.4 & \underline{72.6} \\

MiniCPM-o 4.5~\cite{cui2026minicpmo45}
{\footnotesize\textcolor{gray}{9B}}
& 75.6 & 76.9 & 63.1
& 65.9 & \underline{73.6} & 71.0 \\

\rowcolor{black!8}
\multicolumn{7}{c}{Proprietary MLLMs} \\

Qwen3-VL-Plus~\cite{alibaba2026qwen3vlplus}
& 74.2 & 67.6 & 61.1
& 65.2 & 67.8 & 67.2 \\

Kimi-K2.6~\cite{moonshot2026kimik26}
& 78.0 & 54.4 & 60.7
& 73.9 & 72.6 & 67.9 \\

Gemini-3.5-Flash~\cite{google2026gemini35flash}
& \underline{78.9} & 67.3 & 60.5
& \textbf{81.6} & 71.6 & 72.0 \\

GPT-5.6-Luna~\cite{openai2026gpt56luna}
& 78.6 & 70.1 & 54.4
& 63.8 & 59.6 & 65.3 \\

\rowcolor{black!8}
\multicolumn{7}{c}{Emotion-Oriented MLLMs} \\

EmObserver~\cite{wu2026emobserver}
{\footnotesize\textcolor{gray}{8B}}
& 75.4 & 63.4 & 38.2
& 63.8 & 57.6 & 59.7 \\

OneEmo~\cite{huang2026oneemo}
{\footnotesize\textcolor{gray}{4B}}
& 67.4 & 68.3 & 47.6
& 55.8 & 65.4 & 60.9 \\

\rowcolor{blue!10}
\textbf{UEmo-D1 (Ours)}
{\footnotesize\textcolor{gray}{3B}}
& \textbf{95.4}
& \textbf{86.5}
& \textbf{78.9}
& 72.5
& \textbf{74.0}
& \textbf{81.5} \\

\bottomrule
\end{tabular}%
}
\vskip -0.10in
\end{table*}

\subsection{Detailed Comparison of Recognition Accuracy}

Complementing \cref{fig:5}, Tables~\ref{tab:a4}--\ref{tab:a11} provide detailed comparisons across individual subsets. We report decision-mode results for open-source and emotion-oriented models, as this mode yields better recognition performance in most comparisons, while proprietary models are evaluated in native reasoning mode.

For FER, \cref{tab:a4} covers both static images and dynamic facial expressions. UEmo-D1 achieves an overall accuracy of 66.8\%, exceeding the runner-up, OneEmo, by 10.6\%. This advantage spans both settings: it leads on the static-image subsets CAER-S and FER2013 and outperforms the strongest baseline on the video-based DFEW by 10.2\%. On FERPlus, it remains within 0.2\% of the best result, showing consistently competitive performance across different forms of facial input.

For AER, \cref{tab:a5} reveals substantial variation across models. Emotion-oriented specialization does not consistently outperform general-purpose models, as illustrated by OneEmo's weak performance on RAVDESS, particularly on speech. UEmo-D1, by contrast, leads on both RAVDESS subsets and IEMOCAP and falls only 0.4\% below the best result on CREMA-D. Its overall accuracy of 72.8\% surpasses Qwen2.5-Omni by 18.4\%, representing the largest task-level improvement among the six MER tasks.

The VCEC results are divided into two tables. \Cref{tab:a6} presents MVEI and WebEmo, which assess emotion-statement judgment and hierarchical emotion recognition, respectively. UEmo-D1 leads on three MVEI subtasks and trails the best SP result by only 0.5\%, demonstrating strong performance in judging emotional context and subjectivity. On WebEmo, its margin over the strongest competitor increases from $-2.0$\% to $+9.0$\% and $+9.2$\% across the 2-, 7-, and 25-class settings, highlighting its advantage as recognition becomes more fine-grained. \Cref{tab:a7} reveals a more mixed picture across visual domains. Strong results on EmoSet and both FI subsets contribute to an overall accuracy of 72.9\%, exceeding EmObserver by 4.8\%. However, UEmo-D1 does not lead on Abstract, ArtPhoto, or UnbiasedEmo. These exceptions identify distinctive visual styles and evaluation distributions as areas where broader training coverage may prove valuable.

The CMER results are also presented in two tables. On CMU-MOSEI (\cref{tab:a8}), UEmo-D1 ranks first in two emotion categories and second in three, with no category falling more than 0.4\% below the best result. Its advantage is clearer on sentiment recognition, where it exceeds the strongest baseline by 4.4\%. This contrast becomes more pronounced on MELD (\cref{tab:a9}), with gains of 13.4\% and 10.8\% in emotion and sentiment recognition, respectively, suggesting more effective use of multimodal cues in conversational settings. Together, these gains yield an overall CMER accuracy of 77.6\%, compared with 71.7\% for the strongest baseline, Qwen3-Omni.

Finally, \cref{tab:a10,tab:a11} extend the comparison to ITSA and MSU. UEmo-D1 reaches overall accuracies of 81.8\% and 81.5\%, improving over the strongest baselines by 17.3\% and 8.9\%, respectively, and leads on every subset except MUStARD. These tasks require combining diverse multimodal cues to resolve sentiment, communicative intent, and social meaning. The breadth of these gains suggests that UEmo-D1's strengths extend beyond recognizing individual emotional signals to integrating evidence into higher-level judgments, although the exception on MUStARD leaves room for improvement in interpreting conversational sarcasm.

\begin{figure*}[t]
    \centering
    \includegraphics[width=0.92\linewidth]{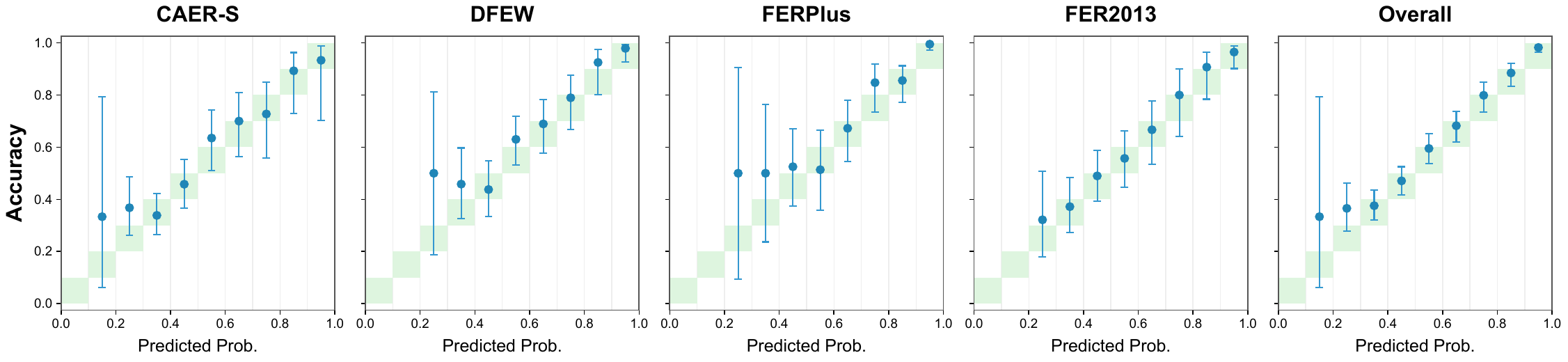}
    \vskip -0.10in
    \caption{Visualization of UEmo-D1's confidence estimation on FER subsets. Green squares mark the perfectly calibrated bins. Blue dots indicate empirical accuracy within each bin, with error bars showing 95\% Wilson confidence intervals.}
    \label{fig:a1}
    \vskip -0.10in
\end{figure*}

\begin{figure*}[t]
    \centering
    \includegraphics[width=0.92\linewidth]{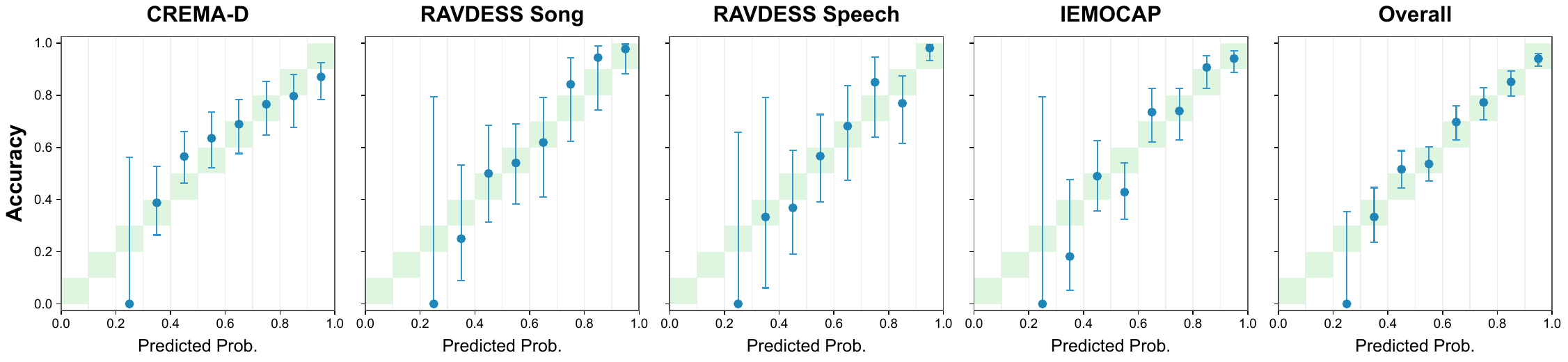}
    \vskip -0.10in
    \caption{Visualization of UEmo-D1's confidence estimation on AER subsets.}
    \label{fig:a2}
    \vskip -0.10in
\end{figure*}

\begin{figure*}[t]
    \centering
    \includegraphics[width=0.92\linewidth]{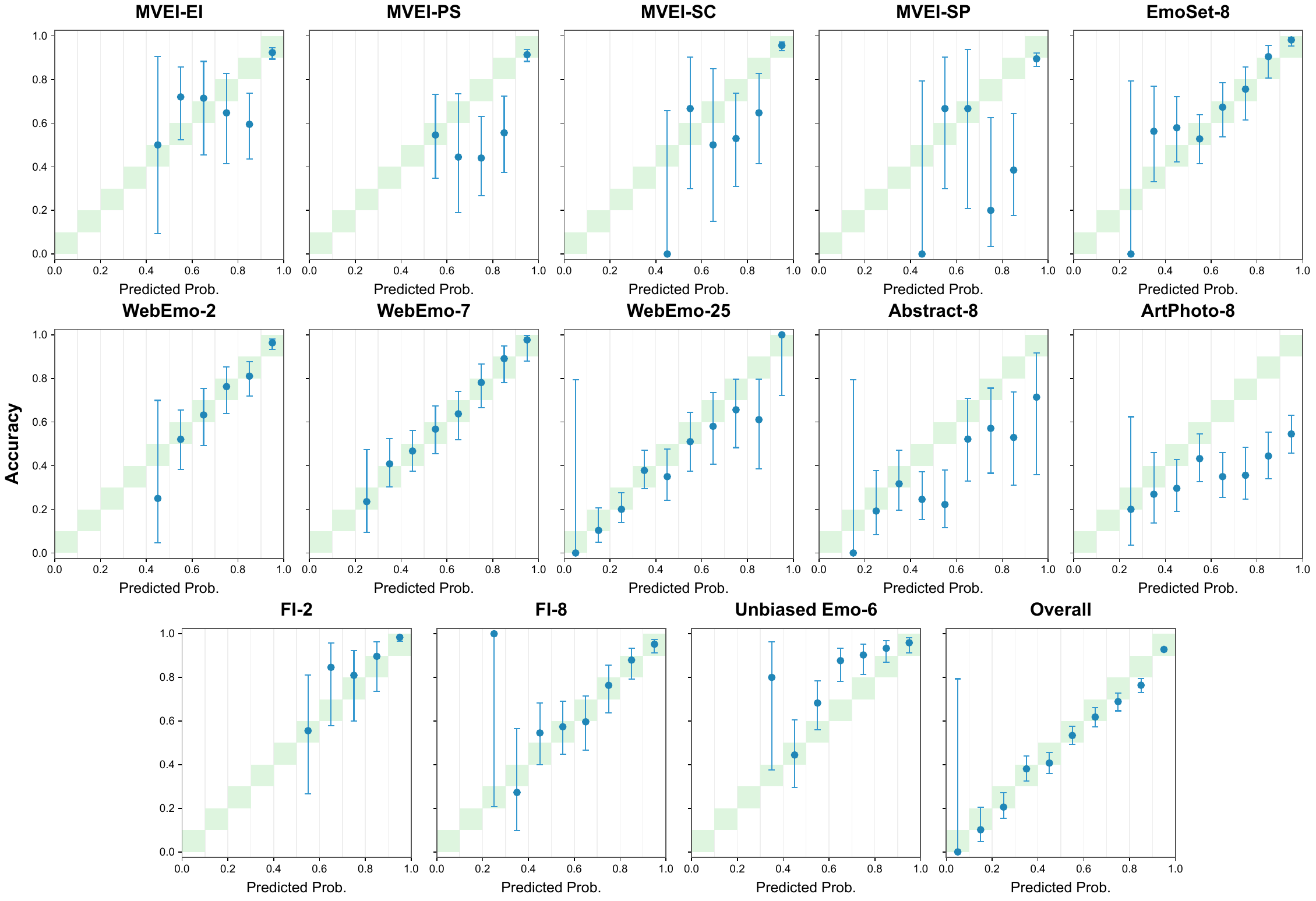}
    \vskip -0.10in
    \caption{Visualization of UEmo-D1's confidence estimation on VCEC subsets.}
    \label{fig:a3}
    \vskip -0.10in
\end{figure*}

\begin{figure*}[t]
    \centering
    \includegraphics[width=1\linewidth]{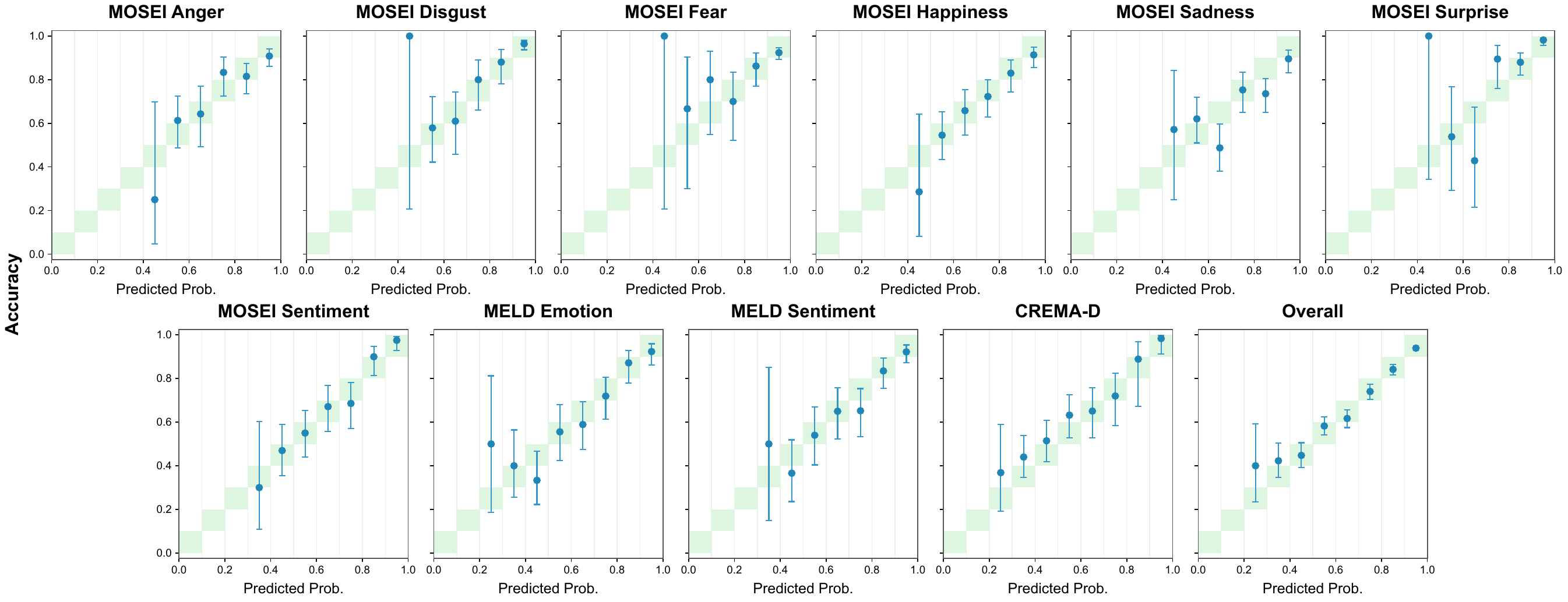}
    \vskip -0.10in
    \caption{Visualization of UEmo-D1's confidence estimation on CMER subsets.}
    \label{fig:a4}
    \vskip -0.10in
\end{figure*}

\begin{figure*}[t]
    \centering
    \includegraphics[width=1\linewidth]{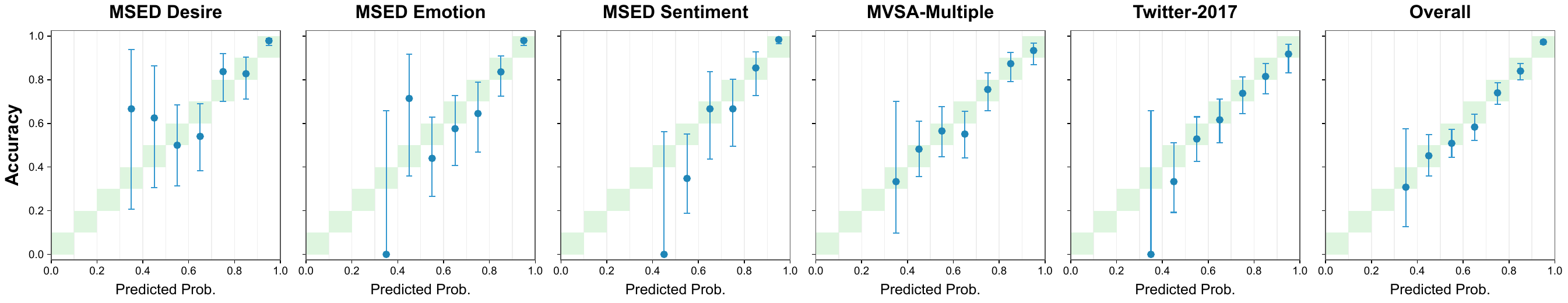}
    \vskip -0.10in
    \caption{Visualization of UEmo-D1's confidence estimation on ITSA subsets.}
    \label{fig:a5}
    \vskip -0.10in
\end{figure*}

\begin{figure*}[t]
    \centering
    \includegraphics[width=1\linewidth]{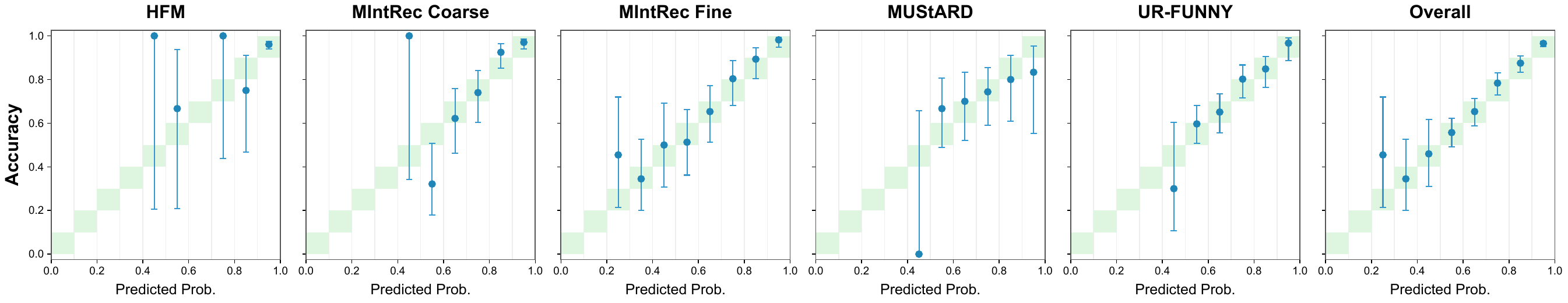}
    \vskip -0.10in
    \caption{Visualization of UEmo-D1's confidence estimation on MSU subsets.}
    \label{fig:a6}
    \vskip -0.10in
\end{figure*}

\subsection{Visualization of Confidence Calibration}

While the main paper focuses on aggregate quantitative comparisons of confidence estimation, Figures~\ref{fig:a1}--\ref{fig:a6} provide complementary visualizations of UEmo-D1 across all subtasks. For FER (\cref{fig:a1}), empirical accuracy generally follows the ideal calibration range at higher predicted confidence levels ($\geq 0.5$), whereas larger deviations appear in lower-confidence bins, a pattern also visible in the pooled results. Several AER subsets (\cref{fig:a2}) exhibit similar variability at low confidence, with sparse samples contributing to wider confidence intervals. These results support the reliability of UEmo-D1's more confident decisions, while identifying low-confidence predictions as an area for further calibration improvement.

For VCEC (\cref{fig:a3}), the pooled results show generally good alignment between confidence and accuracy, although some subsets exhibit larger deviations, including low-confidence bins in FI-8 and Unbiased Emo-6. Abstract and ArtPhoto reveal an additional challenge in visually distinctive domains. Despite a broadly positive relationship between confidence and accuracy, their predictions remain overconfident across much of the confidence range, indicating a need for calibration that better accommodates distribution shifts. The subsequent results for CMER, ITSA, and MSU (\cref{fig:a4,fig:a5,fig:a6}) reveal no additional systematic failure pattern beyond these subset- and confidence-dependent deviations. Taken together, these visualizations support the broad reliability of UEmo-D1's confidence estimates across heterogeneous subtasks, while clarifying where further refinement remains necessary.

\definecolor{fixedgray}{RGB}{100,100,100}

\definecolor{diversegray}{RGB}{100,100,100}

\definecolor{ordergray}{RGB}{100,100,100}

\newcommand{\dplus}[1]{\textcolor{deltagreen}{\ensuremath{+#1}}}
\newcommand{\dminus}[1]{\textcolor{deltared}{\ensuremath{-#1}}}
\newcommand{\dneutral}{\textcolor{gray}{0.0}}

\begin{table*}[t]
\centering
\caption{
Performance of models under different candidate-option orders
on the UEmo-Data test split.
Original uses the same evaluation inputs and option orders as \cref{tab:1}, while Shifted
cyclically moves the first option to the last position,
keeping instructions and multimodal inputs unchanged.
Avg. denotes the unweighted mean across the six tasks,
with uniform-random expected AER accuracy used for EmObserver.
Each $\Delta$ row reports Shifted $-$ Original,
with green indicating improvements and red indicating degradations.
}
\label{tab:a12}
\vskip -0.10in
\setlength{\tabcolsep}{3pt}
\renewcommand{\arraystretch}{1.12}
\small
\resizebox{\textwidth}{!}{%
\begin{tabular}{
lc
*{7}{c>{\color{ordergray}}c}
}
\toprule
\multirow{2}{*}{\textbf{Model}}
& \multirow{2}{*}{\textbf{Mode}}
& \multicolumn{2}{c}{FER (\%)}
& \multicolumn{2}{c}{AER (\%)}
& \multicolumn{2}{c}{VCEC (\%)}
& \multicolumn{2}{c}{CMER (\%)}
& \multicolumn{2}{c}{ITSA (\%)}
& \multicolumn{2}{c}{MSU (\%)}
& \multicolumn{2}{c}{Avg.} \\
\cmidrule(lr){3-4}
\cmidrule(lr){5-6}
\cmidrule(lr){7-8}
\cmidrule(lr){9-10}
\cmidrule(lr){11-12}
\cmidrule(lr){13-14}
\cmidrule(lr){15-16}
& & Original & Shifted
& Original & Shifted
& Original & Shifted
& Original & Shifted
& Original & Shifted
& Original & Shifted
& Original & Shifted \\
\midrule

\rowcolor{black!8}
\multicolumn{16}{c}{Emotion-Oriented MLLMs} \\

AffectGPT~\cite{lian2025affectgpt}
{\footnotesize\textcolor{gray}{7B}}
& D
& 15.2 & 10.4
& 23.0 & 28.1
& 34.3 & 33.4
& 62.4 & 54.3
& 58.1 & 59.7
& 61.9 & 56.8
& 42.5 & 40.5 \\
\quad $\Delta$ &
& \multicolumn{2}{c}{\dminus{4.8}}
& \multicolumn{2}{c}{\dplus{5.1}}
& \multicolumn{2}{c}{\dminus{0.9}}
& \multicolumn{2}{c}{\dminus{8.1}}
& \multicolumn{2}{c}{\dplus{1.6}}
& \multicolumn{2}{c}{\dminus{5.1}}
& \multicolumn{2}{c}{\dminus{2.0}} \\

Nano-EmoX~\cite{huang2026nanoemox}
{\footnotesize\textcolor{gray}{1.5B}}
& D
& 14.4 & 14.9
& 19.0 & 15.6
& 32.4 & 32.8
& 54.6 & 25.2
& 23.9 & 21.0
& 38.6 & 42.1
& 30.5 & 25.3 \\
\quad $\Delta$ &
& \multicolumn{2}{c}{\dplus{0.5}}
& \multicolumn{2}{c}{\dminus{3.4}}
& \multicolumn{2}{c}{\dplus{0.4}}
& \multicolumn{2}{c}{\dminus{29.4}}
& \multicolumn{2}{c}{\dminus{2.9}}
& \multicolumn{2}{c}{\dplus{3.5}}
& \multicolumn{2}{c}{\dminus{5.2}} \\

EmObserver~\cite{wu2026emobserver}
{\footnotesize\textcolor{gray}{8B}}
& D
& 52.6 & 53.7
& -- & --
& 68.1 & 69.3
& 66.5 & 65.6
& 61.0 & 62.4
& 59.7 & 56.8
& 54.3 & 54.3 \\
\quad $\Delta$ &
& \multicolumn{2}{c}{\dplus{1.1}}
& \multicolumn{2}{c}{--}
& \multicolumn{2}{c}{\dplus{1.2}}
& \multicolumn{2}{c}{\dminus{0.9}}
& \multicolumn{2}{c}{\dplus{1.4}}
& \multicolumn{2}{c}{\dminus{2.9}}
& \multicolumn{2}{c}{\dneutral} \\

OneEmo~\cite{huang2026oneemo}
{\footnotesize\textcolor{gray}{4B}}
& D
& 56.2 & 50.1
& 27.5 & 17.3
& 57.2 & 51.3
& 67.3 & 65.7
& 62.6 & 60.0
& 60.9 & 55.1
& 55.3 & 49.9 \\
\quad $\Delta$ &
& \multicolumn{2}{c}{\dminus{6.1}}
& \multicolumn{2}{c}{\dminus{10.2}}
& \multicolumn{2}{c}{\dminus{5.9}}
& \multicolumn{2}{c}{\dminus{1.6}}
& \multicolumn{2}{c}{\dminus{2.6}}
& \multicolumn{2}{c}{\dminus{5.8}}
& \multicolumn{2}{c}{\dminus{5.4}} \\

\rowcolor{blue!10}
\textbf{UEmo-D1 (Ours)}
{\footnotesize\textcolor{gray}{3B}}
& D
& \textbf{66.8} & \textbf{66.5}
& \textbf{72.8} & \textbf{73.4}
& \textbf{72.9} & \textbf{72.6}
& \textbf{77.6} & \textbf{77.5}
& \textbf{81.8} & \textbf{81.9}
& \textbf{81.5} & \textbf{81.5}
& \textbf{75.6} & \textbf{75.6} \\
\rowcolor{blue!10}
\quad $\Delta$ &
& \multicolumn{2}{c}{\dminus{0.3}}
& \multicolumn{2}{c}{\dplus{0.6}}
& \multicolumn{2}{c}{\dminus{0.3}}
& \multicolumn{2}{c}{\dminus{0.1}}
& \multicolumn{2}{c}{\dplus{0.1}}
& \multicolumn{2}{c}{\dneutral}
& \multicolumn{2}{c}{\dneutral} \\

\bottomrule
\end{tabular}%
}
\vskip -0.10in
\end{table*}

\begin{table}[t]
\centering
\caption{
Accuracy comparison on CA-MER under consistent and conflicting
multimodal conditions, with 500 samples per condition.
\textbf{Bold} and \underline{underline} indicate the best and
second-best results in each column, respectively.
}
\label{tab:a13}
\vskip -0.10in
\setlength{\tabcolsep}{4pt}
\renewcommand{\arraystretch}{1.12}
\small
\resizebox{\columnwidth}{!}{%
\begin{tabular}{lcccc}
\toprule
\multirow{3}{*}{\textbf{Model}}
& \multicolumn{4}{c}{CA-MER: Accuracy (\%) $\uparrow$} \\
\cmidrule(lr){2-5}
& \multirow{2}{*}{Consistent}
& \multicolumn{3}{c}{Conflicting} \\
\cmidrule(lr){3-5}
& & Audio-aligned & Video-aligned & Avg. \\
\midrule

\rowcolor{black!8}
\multicolumn{5}{c}{Emotion-Oriented MLLMs} \\

AffectGPT~\cite{lian2025affectgpt}
{\footnotesize\textcolor{gray}{7B}}
& 49.8 & 40.6 & 30.0 & 35.3 \\

Nano-EmoX~\cite{huang2026nanoemox}
{\footnotesize\textcolor{gray}{1.5B}}
& 12.6 & 17.4 & 11.6 & 14.5 \\

OneEmo~\cite{huang2026oneemo}
{\footnotesize\textcolor{gray}{4B}}
& \underline{63.4}
& \underline{46.0}
& \underline{36.8}
& \underline{41.4} \\

\rowcolor{blue!10}
\textbf{UEmo-D1 (Ours)}
{\footnotesize\textcolor{gray}{3B}}
& \textbf{66.2}
& \textbf{54.0}
& \textbf{41.2}
& \textbf{47.6} \\

\bottomrule
\end{tabular}%
}
\vskip -0.10in
\end{table}

\subsection{Other Diagnostic Experiments}

\textbf{Sensitivity to Option Order.}
Emotion-oriented models are typically optimized on specialized data, potentially at the expense of some generalization capabilities. While reduced proficiency in unrelated domains such as mathematics may be less consequential for their intended use, this trade-off becomes more concerning when it affects how candidate answers are presented, as users may arrange them in arbitrary orders. To examine this sensitivity, we cyclically shift the options of every test sample in UEmo-Data while keeping its instruction and multimodal content unchanged. As shown in \cref{tab:a12}, AffectGPT, Nano-EmoX, and OneEmo exhibit average accuracy drops of 2.0--5.4\%, with the largest task-level decline reaching 29.4\% for Nano-EmoX on CMER. Such deterioration under a purely positional change reveals a reliance on option arrangements beyond their semantic content.

Against this backdrop, UEmo-D1 and EmObserver maintain nearly unchanged average accuracy. UEmo-D1's more consistent behavior may partly stem from option-order randomization during training, illustrating the value of a simple measure against positional shortcuts. EmObserver's emphasis on input diversity may offer a related benefit, helping prevent the learned decision space from collapsing into rigid associations between semantic labels and option positions.

\textbf{Influence of Conflicting Modalities.}
CA-MER~\citeapp{han2025benchmarking} is designed to evaluate emotion recognition under modality conflicts through three conditions: consistent, where both modalities agree with the ground-truth emotion, and audio-aligned and video-aligned, where only the corresponding modality reflects it. As shown in \cref{tab:a13}, UEmo-D1 outperforms the compared models across all three conditions, but its performance also reveals room for improvement. In particular, accuracy reaches 54.0\% on audio-aligned samples but falls to 41.2\% on video-aligned samples. This asymmetry is consistent with the audio-dominant tendency reported in the original CA-MER study, suggesting that UEmo-D1 may likewise rely more heavily on acoustic evidence when modalities disagree. Thus, despite its overall advantages, resolving conflicting cues remains a challenge, motivating more targeted training on modality conflicts to encourage balanced use of audio and visual evidence.

\begingroup
    \small
    \bibliographystyleapp{ieeenat_fullname}
    \bibliographyapp{main}
    \par
    \clearpage
\endgroup

\end{document}